\documentclass[journal]{IEEEtran}
\usepackage{amsmath}
\usepackage{graphicx} 
\usepackage{subcaption}
\usepackage{float}
\usepackage{multirow}
\usepackage{balance}
\usepackage{cite}
\usepackage[
  colorlinks=true,
  citecolor=blue,
  linkcolor=black,
  urlcolor=blue
]{hyperref}
\ifCLASSINFOpdf
\else
\fi
\title{Performance Enhancement of Gas Electron Multipliers Using an Optimized Single-Conical Hole Geometry for Different Charged Particles}
\author{Poojan Angiras$^{1}$,
        Sachin Rana$^{1}$,
        Md. Kaosor Ali Mondal$^{1}$,
        Shakir Eqbal$^{1}$,
        Amal Sarkar$^{1}$ \\
        $^{1}$ School of Physical Sciences, Indian Institute of Technology Mandi, Kamand, Mandi - 175005, India \\
        poojan.angiras@cern.ch, sachin.rana@cern.ch, mohommad.mondal@cern.ch, di2505@students.iitmandi.ac.in, amal.sarkar@cern.ch
}        

\begin{document}
\bstctlcite{BSTcontrol}
\maketitle
\begin{abstract}
Gas Electron Multipliers (GEMs) are essential detector components in modern high-energy physics experiments, where precise and stable detection of charged particles over a broad energy range is required. We present a comprehensive Garfield$^{++}$ and ANSYS-based study of conventional bi-conical and optimized single-conical GEM detectors to investigate the performance of the GEM detector for muons ($\mu$), pions ($\pi$), kaons ($K$), and protons ($P$), which constitute the dominant charged particles measured directly in collider-based experiments. The aim is to examine the impact of particle-dependent ionization characteristics on charge amplification and ion backflow, and to evaluate the potential of an optimized GEM configuration for different leptons and hadrons. The conventional bi-conical GEM design does not always operate at optimal efficiency, as ion backflow can lead to space-charge accumulation and electric field distortions, ultimately limiting performance in high-rate environments. Thus, geometrical optimization is essential to address these limitations and enhance detector performance. A single-conical hole geometry is introduced and systematically compared with the conventional bi-conical design. For both of these configurations, the results exhibit clear and systematic variations in the detector performance with the particle type and the incident energy. The optimized geometry improves the balance between effective gain and ion backflow, demonstrating its potential for future high-rate MPGD applications.

%The study demonstrates that the optimized GEM geometry provides a better balance between charge amplification and ion suppression while maintaining stable operation in high-rate environments, highlighting its potential for future GEM-based detectors in high-energy physics experiments.
\end{abstract}
%Captions: Effective gain as a function of the incident particle energy for muons, pions, kaons, and protons in the standard bi-conical/optimized single-conical GEM detector. The points represent the simulated effective gain, while the lines are included to guide the eye. Error bars denote the statistical uncertainties obtained from the Garfield$^{++}$ simulations.

\section{Introduction}

Gaseous detectors have been used for more than a century in experimental physics to detect charged particles. Early detectors such as ionization chambers, Geiger–Müller counters, and multi-wire proportional chambers enabled precise measurements of particle flux, position, and energy deposition \cite{MPGDReview1999}. %Although MWPCs significantly advanced particle tracking, their performance is limited in high-rate environments due to space-charge effects, detector aging, and restricted rate capability.
MWPCs have made considerable progress in particle tracking; however, their performance is limited by space-charge effects, detector aging, and limited high-rate capabilities. Modern high-energy physics experiments are producing ever-higher luminosities, which in turn require detectors capable of operating at high particle fluxes while preserving high spatial resolution~\cite{Kudryavtsev2017}, timing accuracy, and long-term stability~\cite{Malhotra2018}. This led to the creation of Micro-Patterned Gas Detectors (MPGDs), which use precisely patterned electrode structures to achieve efficient charge amplification \cite{MSGCProblems, Buzulutskov2007}. Technologies such as Micromegas and Gas Electron Multipliers (GEMs) can surpass conventional gaseous detectors in terms of rate capability, ion feedback, and signal quality~\cite{CHARPAK200226}.

One of the most popular gas detector technologies among MPGDs is the Gas Electron Multiplier (GEM), originally proposed by Sauli in 1997 \cite{Sauli1997GEM}. A GEM consists of a Kapton thin foil coated on both sides with a thin conductive layer and pierced by a high-density pattern of microscopic holes~\cite{Sauli2016Review}. The high electric field in these holes causes the electrons created in the drift region to undergo avalanche multiplication, and several GEM foils can be stacked to obtain high gain with stable operation\cite{micropattern}. These characteristics make GEM detectors widely used in tracking systems, time projection chambers, and muon detectors \cite{GEMApplications2003}.

A fundamental difficulty in GEM detectors is ion backflow (IBF), in which some of the positive ions generated during avalanche multiplication wander back into the drift region, altering the electric field and reducing detector performance, especially in continuous-readout systems \cite{Tarafdar2022, Sauli2006}. A major objective is therefore to reduce the ion backflow while maintaining a high effective gain. Common approaches include optimizing the electric-field configuration, employing multi-stage GEM structures, and modifying the GEM hole geometry \cite{Bouianov2001}.

The shape of GEM holes has a considerable influence on the electric field, electron collection, charge amplification, and ion transport~\cite{Keller_2020}. Since these processes govern the detector's overall performance, optimizing the microscopic hole geometry has emerged as an effective approach to improving GEM performance without altering the detector's overall architecture. In recent years, several studies have explored alternative hole geometries to enhance electron transparency, improve charge multiplication, and suppress ion backflow, highlighting the strong dependence of detector performance on the hole profile~\cite{Mondal_2025,Bhattacharya2025}. Such differences in the hole geometries influence the detector response under realistic operating conditions and have therefore become an important research field for the development of next-generation gaseous detectors.

This optimization relies on electric-field calculations using ANSYS Mechanical APDL~\cite{Martyanov2014} together with microscopic charge-transport calculations performed with Garfield$^{++}$~\cite{GarfieldGeneral}. These tools accurately describe electron and ion transport, gas ionization, diffusion, and avalanche formation, allowing systematic investigation of the detector response under different operating conditions~\cite{Jagielski2023, Veenhof1998}.

In high-energy physics experiments, GEM detectors are exposed to a wide range of charged particles, including electrons, muons, pions, kaons, and protons.  The differences in their mass, energy loss, and ionization properties lead to differences in the primary electron production, which directly affect detector gain and ion backflow. Understanding such effects is crucial to optimizing detector performance under various experimental conditions.

%In this work, we present a comprehensive study of GEM detector responses to different charged particles by comparing conventional and optimized hole geometries. The electric field distributions are calculated with ANSYS Mechanical APDL and interfaced with Garfield$^{++}$ to model charge transport, avalanche multiplication, and ion dynamics. Primary ionization is modeled using a Bethe--Bloch-based approach, and the dependence of effective gain and ion backflow on particle type and incident energy is systematically investigated for both single- and triple-GEM configurations. The influence of hole geometry on detector performance is analyzed with particular emphasis on achieving an improved balance between charge amplification and ion suppression for future high-rate gaseous detectors.

%%In this work, we present a comprehensive investigation of GEM detector performance through a systematic comparison of conventional bi-conical and optimized single-conical hole geometries. By examining the detector response to different charged particles over a wide range of incident energies, we identify the influence of hole geometry on charge amplification and ion suppression, providing important insights for the optimization of next-generation high-rate gaseous detectors.

In this work, we present a comprehensive investigation of the impact of GEM hole geometry on detector performance by systematically comparing conventional bi-conical and optimized single-conical designs. The detector response to muons, pions, kaons, and protons is examined over a broad energy range to quantify the influence of hole geometry on effective gain and ion backflow. The results establish optimized hole geometry as an effective design parameter for achieving simultaneous enhancement of charge amplification and suppression of ion backflow in high-rate gaseous detectors. Realistic electric-field maps are generated using ANSYS Mechanical APDL and imported into Garfield$^{++}$ to model primary ionization, electron transport, avalanche multiplication, and ion drift. Primary ionization is described using the Bethe-Bloch formalism, and the dependence of the effective gain and ion backflow on particle species and incident energy is systematically evaluated for both single and triple-GEM configurations. The influence of hole geometry on detector performance is evaluated, with particular emphasis on enhancing effective gain while suppressing ion backflow, providing guidance for the development of next-generation high-rate gaseous detectors. Unlike previous studies, which mainly focused on electron transport or detector optimization under fixed irradiation conditions, the present work systematically investigates the response of conventional and optimized GEM geometries to multiple charged particle species over a broad energy range, providing a unified understanding of geometry-dependent charge amplification and ion suppression relevant to future high-rate collider detectors.

\section{working principle}
The operation of a triple Gas Electron Multiplier (GEM) detector is based on the sequential processes of primary ionization, electron transport, avalanche multiplication, and signal induction within a finely structured gaseous medium. A charged particle loses energy predominantly through ionization as described by the Bethe–Bloch formalism~\cite{PDGBetheBloch}, when it passes through the drift region filled with an Ar:CO$_2$ (70:30) gas mixture. This results in the formation of primary electron-ion pairs along its path. %$, such as electrons (e), muons ($\mu$), pions ($\pi$), kaons (K), and protons (p)

  \begin{figure}[!t]
  \centering
\includegraphics[
    width=1.05\columnwidth,
    height=0.35\textheight,
    keepaspectratio
]{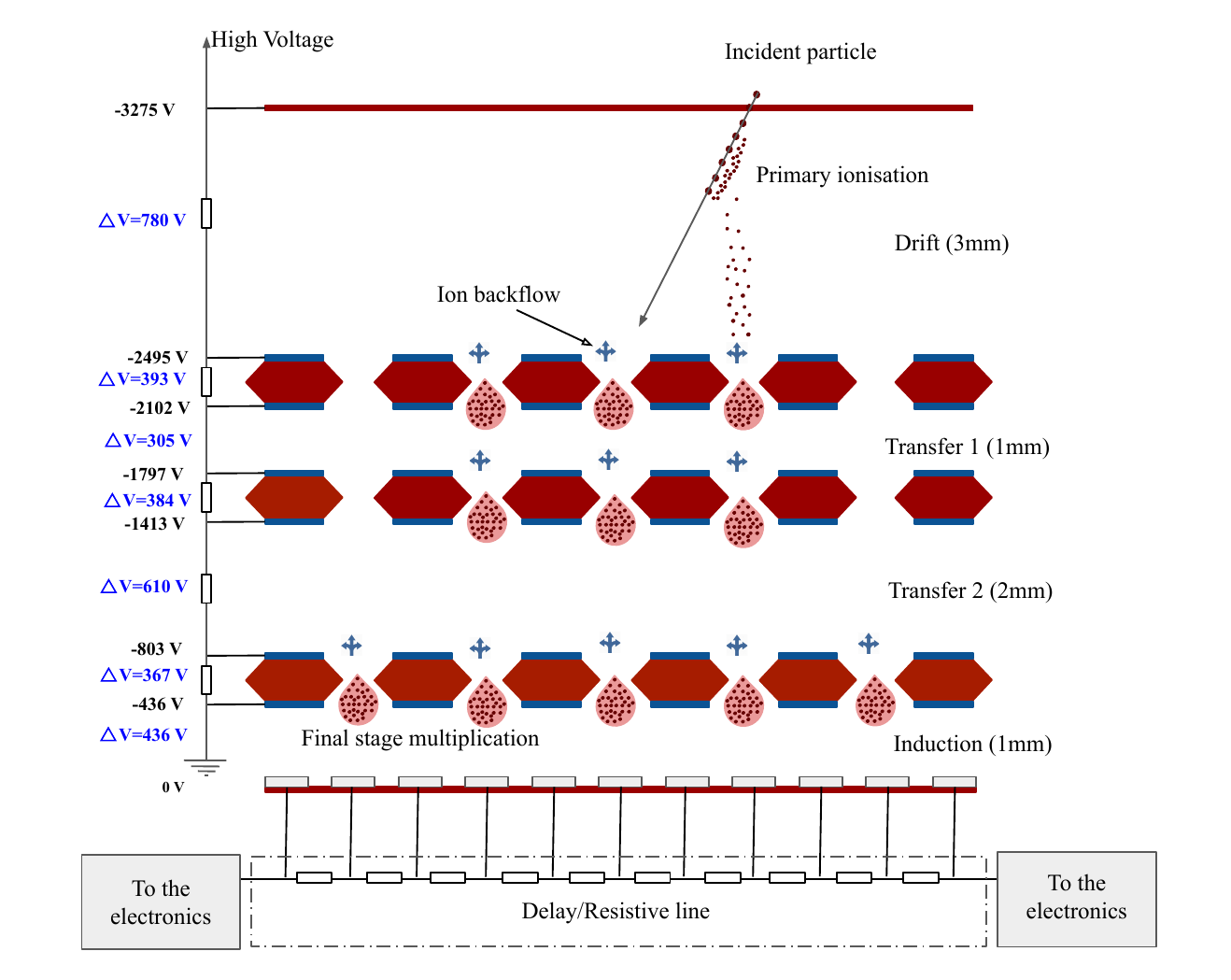}  \caption{Schematic of the triple-GEM detector geometry showing the drift, transfer, and induction regions, together with the applied electrode potentials and voltage differences across each region. The primary ionization process, successive electron avalanche multiplication, ion backflow, and the readout electronics are also shown.}
  \label{fig:STG}
\end{figure}
These electrons are guided toward the first GEM foil under the influence of a uniform electric field in the drift region. Upon entering the microscopic holes of the first GEM foil, where the electric field reaches several tens of kV/cm, the electrons undergo avalanche multiplication through ionizing collisions. The resulting amplified electron cloud is extracted into the first transfer region and efficiently transported to the second GEM foil, where the multiplication process is repeated as shown in Fig.~\ref {fig:STG}.

This staged amplification continues through the third GEM foil, resulting in an exponential increase in the number of charge carriers and achieving an overall detector gain~\cite{AbiAkl2016}. Following the final amplification stage, the electrons are collected at the readout anode, producing the detector signal. Positive ions are produced during the multiplication drift in the opposite direction, most collected by the GEM electrodes, while a small fraction returns to the drift region, causing ion backflow~\cite{Chatterjee2026}. The multi-layer GEM configuration plays a crucial role in suppressing this ion feedback by effectively trapping ions at intermediate stages while preserving high electron transmission and gain~\cite{Bouclier1997IEEE}. The overall detector response is therefore governed by the electric-field configuration, the gas properties, and detector geometry~\cite{PATRA201725}.
\section{Detector Design and Computational Framework}

\subsection{Standard GEM detector Geometry}

The conventional GEM foil design widely employed in high-energy physics experiments is used as the reference geometry in this work. It consists of a 50~$\mu$m thick polyimide (Kapton) substrate coated on both sides with 5~$\mu$m thick copper layers. The foil is perforated with a regular hexagonal array of microscopic holes having a pitch of 140~$\mu$m. In the conventional bi-conical geometry, each hole has a minimum diameter of approximately 50~$\mu$m at the center of the foil and expands symmetrically to an outer diameter of 70~$\mu$m at both the top and bottom copper layers. This bi-conical hole profile is typically fabricated using the well-established double-mask chemical etching technique and has become the standard GEM geometry due to its reliable performance and mechanical stability~\cite{Bouhali2022}. In this work, it serves as the reference geometry for evaluating the performance of the optimized hole design.

\subsection{Hole Geometry Optimization}
The performance of a GEM detector is strongly influenced by the geometry of its microscopic holes, which directly affects electron and ion transport, avalanche multiplication, and ion collection via the electric-field distribution within the holes. Therefore, modifying the hole profile provides a simple and effective way to improve detector performance without changing the overall detector configuration. To investigate this effect, the conventional bi-conical geometry is compared with the optimized single-conical geometry.

\begin{figure}[!t]
    \centering
    \includegraphics[width=0.75\columnwidth]{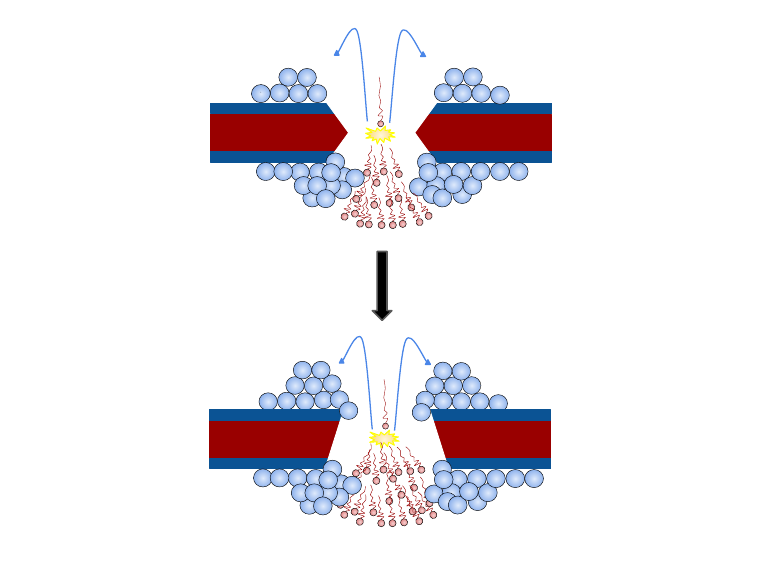}
    \caption{Schematic illustration of the ion collection mechanism in the conventional bi-conical (top) and optimized single-conical (bottom) GEM hole geometries. The red lines represent the electron trajectories, while the blue spheres represent the positive ions produced during avalanche multiplication.}
    \label{fig:modified_hole}
\end{figure}

In the conventional bi-conical geometry, the avalanche develops near the middle of the hole, where the upper copper electrode is relatively farther from the multiplication region. Therefore, the positive ions produced during the avalanche have to travel a longer distance before being collected, allowing a fraction of them to drift back into the drift region and contribute to ion backflow. In contrast, the asymmetric profile of the proposed single-conical geometry brings the upper copper electrode closer to the avalanche region, as illustrated in Fig.~\ref{fig:modified_hole}. As a result, the positive ions have a shorter path to the electrode and are more likely to be collected before escaping into the drift region. This increases ion-collection efficiency at the upper electrode, thereby reducing ion backflow while preserving electron transport and the overall detector architecture.

Based on this concept, the conventional bi-conical hole geometry was replaced with an optimized single-conical profile having a 50~$\mu$m upper opening and a 70~$\mu$m lower opening. An additional practical advantage of the proposed geometry is that it can be fabricated using the single-mask etching technique, which simplifies the fabrication process compared with the conventional double-mask process used for bi-conical GEM foils. In addition to the hole profile, all other detector parameters, such as foil thickness, hole pitch, materials, detector dimensions, gas mixture, and operating voltages, were kept unchanged. This ensures that any changes observed in the detector response arise solely from the modified hole geometry.

\subsection{Detector Configuration and Gas Parameters}
The GEM foil is placed in a simplified detector setup which consists of three regions: a drift region above the GEM, an amplification region within the GEM holes, and an induction region below the GEM. The drift region guides the primary electrons into the GEM holes and the induction region collects the multiplied charge on the readout electrode. For this study, typical dimensions are used with a drift gap of 3 mm and an induction gap of about 1 mm, as as shown in Fig~\ref{fig:STG}. Transfer gaps of 1 mm and 2 mm are used for the first and second transfer regions, respectively.

The gas mixture is an important factor in determining the electron transport properties, ion mobility, and avalanche characteristics. In the present work, a common mixture of Argon and Carbon Dioxide (Ar/CO${_2}$ in a 70:30 ratio) is used due to its favorable qualities, such as good electron drift velocity, moderate diffusion, and steady operation under high fields~\cite{BACHMANN2002294}. Typical field values are selected to obtain efficient electron transport and amplification. Drift fields are of the order of 1–3 kV/cm, induction fields of the order of 4–5 kV/cm and a high electric field in the GEM holes is generated by a voltage difference across the foil~\cite{AlAtoum2020}.  Both geometries use the same parameters to essentially isolate the effect of the hole shape on the detector performance~\cite{Hilke2020}.

\subsection{Electric Field Map Preparation using ANSYS Mechanical APDL}
The electric field distribution inside the GEM structure is estimated using ANSYS Mechanical APDL\cite{ANSYS2023}, which provides a finite element method (FEM) based solution of electrostatic fields in complex geometries.
  \begin{figure}[!t]
  \centering
  \includegraphics[width=\columnwidth]{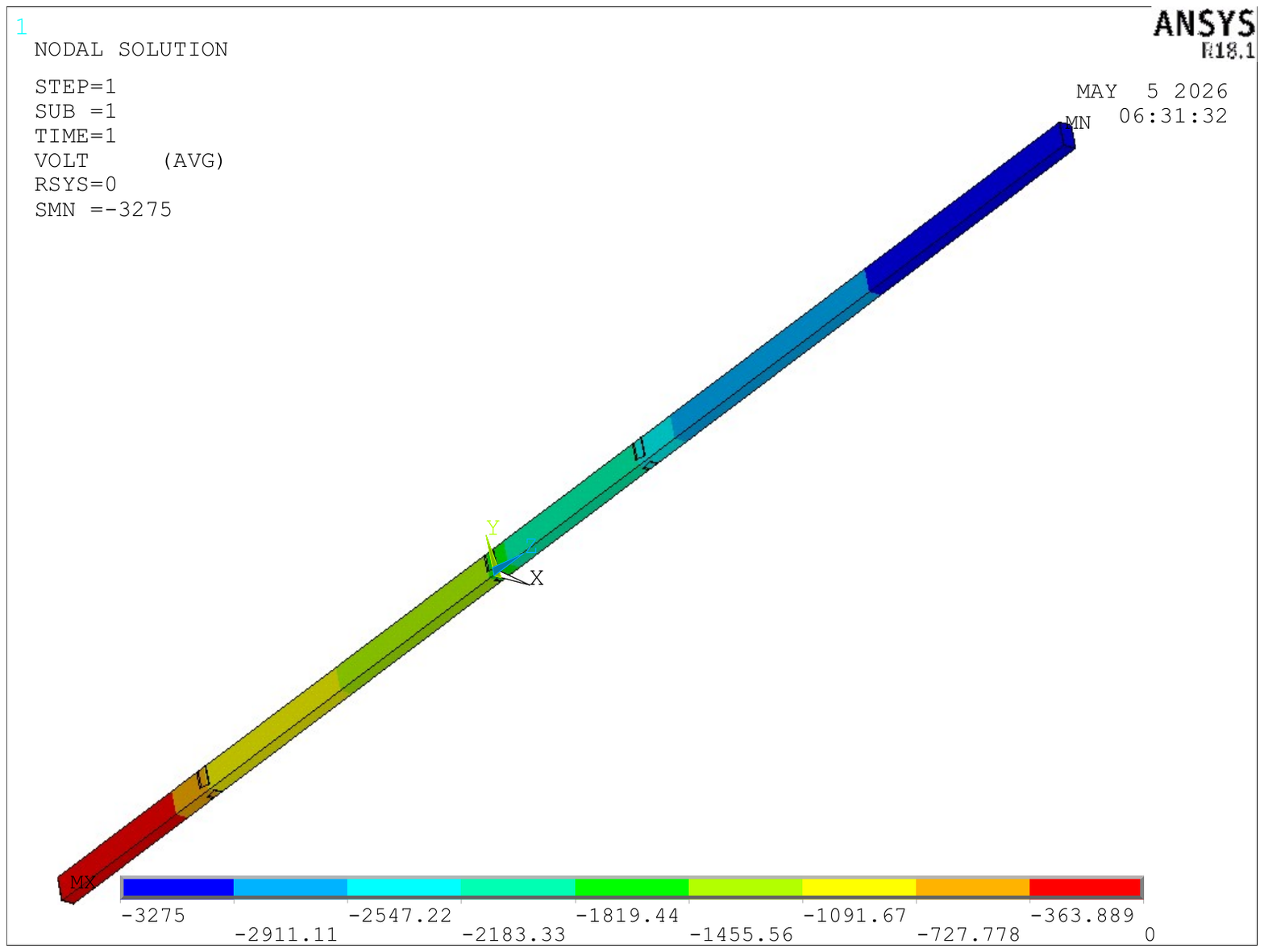}
  \caption{Detector geometry designed in ANSYS Mechanical APDL using finite-element analysis. The geometry includes the drift, GEM, transfer, and induction regions with their corresponding applied voltages. The resulting electric-field maps are then imported into Garfield$^{++}$ for the study of electron transport, avalanche multiplication, and ion backflow.}
  \label{fig:APDL}
\end{figure}
The detailed GEM geometry, including the Kapton substrate, copper layers, and hole structure is constructed in ANSYS environment. Appropriate boundary conditions are applied by applying fixed potentials to the copper electrodes to establish the electric field across the GEM as shown in Fig.~\ref{fig:APDL} \cite{BHATTACHARYA201764}.

A fine mesh is employed in the vicinity of the GEM holes to accurately capture the strong field gradients that drive avalanche multiplication. Particular attention is given to resolving the field inside the hole, where the electric field can reach values on the order of tens of kV/cm. The resulting 3D electric field maps are then exported and formatted to be compatiblle with Garfield$^{++}$, providing a seamless interface between field computation and particle transport studies. This ensures that the microscopic processes are based on realistic, high-resolution electric field distributions.

\section{Computational Methodology}
A detailed study has been carried out to examine the response of Gas Electron Multiplier (GEM) detectors to various charged-particle species.  The methodology combines realistic particle generation, modeling of primary ionization, microscopic charge transport, and statistical analysis of detector observables. The goal is to obtain a predicted detector response that correctly describes the physical processes controlling the operation of the GEM detector under different conditions.
\subsection{Incident Particle Configuration}
The detector response is studied for a representative set of charged particles, commonly observed in high-energy physics experiments, namely muons ($\mu$), pions ($\pi$), kaons ($\kappa$), and protons ($p$). These particles span a large range in mass and interaction properties and are therefore useful to test the dependence of detector performance on particle type. The detector response is studied for each particle species at incident energies up to 10 GeV, which is the energy range of interest for high-energy physics experiments.
Charged particles were injected normal to the detector plane and traversed the entire drift gap, ensuring full primary ionization.

The incident particle goes straight through the drift region perpendicular to the GEM surface.
This setup reduces the geometric complexity while keeping the basic physics of particle–gas interactions. The length of the track is chosen to be large enough to cover the whole drift gap and therefore to pick up the whole ionization profile of each particle. The same incident geometry is maintained for all particle types so that any observed variation in the detector response can be directly attributed to the differences in the particle properties.

\subsection{Primary Ionization}

Primary electron-ion pairs are produced when charged particles pass through the detector gas. These primary electrons provide the starting point for electron transport and avalanche multiplication inside the GEM detector. In this work, the energy loss of charged particles in the gas is described using the Bethe-Bloch formalism, which estimates the number of ionization clusters produced for different particle types and incident energies.

For each particle track, the deposited energy is converted into ionization clusters, each producing a number of primary electrons. The positions of these clusters along the particle track are statistically generated to reflect the stochastic nature of the ionization process. Because different particles deposit different amounts of energy in the gas, the number of primary electrons also varies with the particle type and energy. The generated primary electrons are then used as the input to subsequent electron transport and avalanche multiplication.

\subsection{Electron Transport and Avalanche Process}
The electric-field maps generated in ANSYS Mechanical APDL are imported into Garfield$^{++}$, where the transport of charged particles and avalanche multiplication within the detector are investigated. The generated primary electrons are transported through the drift region under the applied electric field. The transport process is studied using microscopic tracking in Garfield$^{++}$, which accounts for the velocity of electron drift, transverse and longitudinal diffusion, and interactions with gas molecules~\cite{Amoroso2024}. The electron trajectories are influenced by the detailed electric field maps imported from ANSYS, ensuring that field non-uniformities near the GEM holes are properly captured.

\begin{figure}[!t]
    \centering
    \includegraphics[width=0.8\columnwidth]{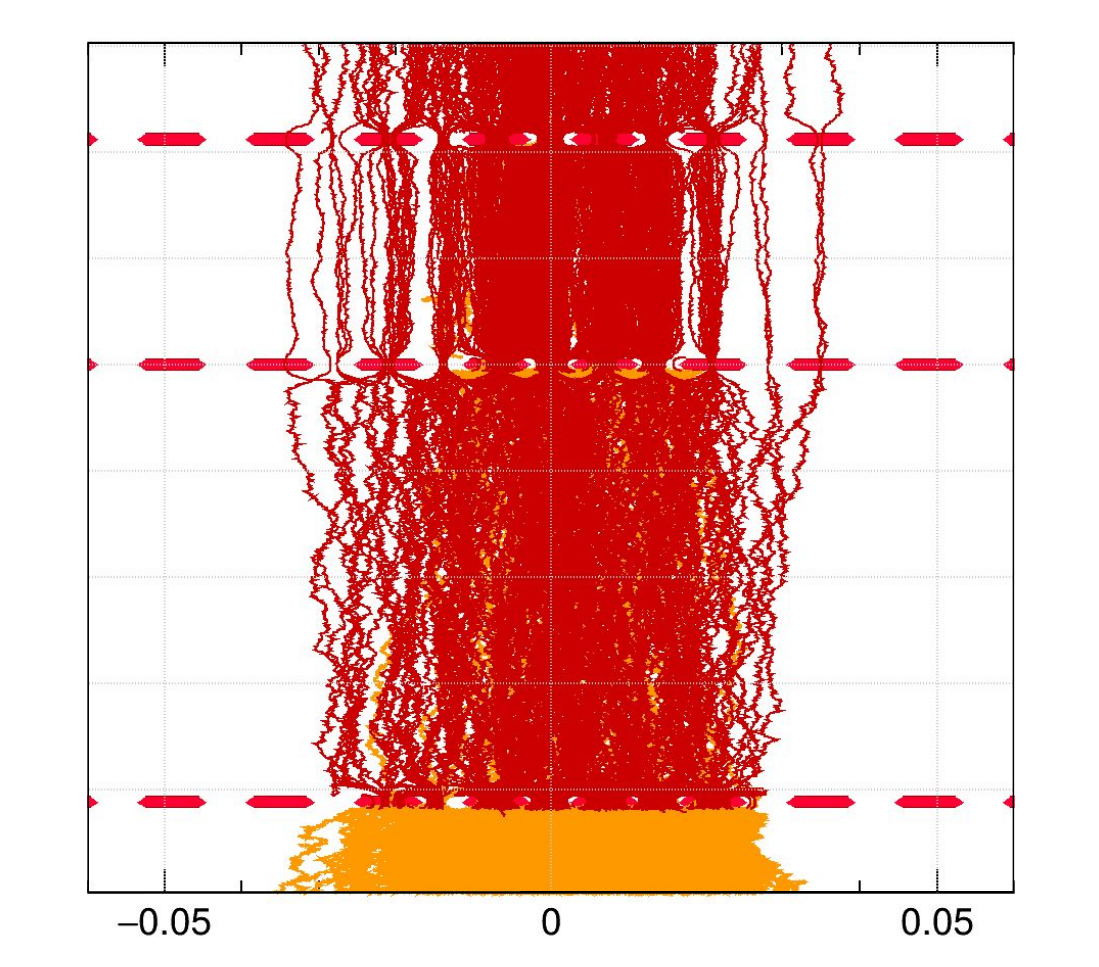}
    \caption{Electron avalanche in Garfield$^{++}$, where the red lines denote the trajectories of positive ions drifting back towards the drift region, while the yellow lines show the trajectories of electrons.}
    \label{fig:Garfield}
\end{figure}
As electrons enter the high-field region inside the GEM holes, they undergo avalanche multiplication through successive ionizing collisions~\cite{Bachmann1999}. The trajectories of avalanche electrons and back-drifting positive ions are illustrated in Fig~\ref{fig:Garfield}, where yellow and red lines represent the path of electron and ion-backflow, respectively. This process is governed by the Townsend ionization coefficient which depends on the local electric field and the gas composition. Garfield$^{++}$ tracks the path and multiplication history of each electron on the microscopic scale, enabling a detailed description of the avalanche process~\cite{Bonivento2002}. The total number of electrons produced in the avalanche is recorded for each primary electron and used to calculate the effective gain~\cite{Bouhali2018}.

In addition to the electron transport the motion of the positive ions generated by the avalanche is considered. Positive ions move much more slowly than electrons because they are heavier. Their trajectories are tracked until they are collected on the GEM electrodes or drift back into the drift region and contribute to the ion backflow (IBF).
 
\subsection{Collection of Charge and Signal Formation}
The detector signal is determined by the collection of multiplied electrons at the readout electrode.  For each event, the total number of electrons arriving in the induction region is recorded. This parameter is directly related to the detector's effective gain and represents the combined effects of electron transit, avalanche multiplication, and extraction efficiency. The corresponding positive ions produced during avalanche multiplication are also tracked, and the ion backflow fraction is determined from the fraction of ions that drift back into the drift region. This requires careful tracking of ion trajectories over relatively long timescales compared to electron motion. The balance between electron collection and ion backflow give a complete picture of detector performance, particularly under high-rate conditions, where ion buildup can strongly influence detector operation.
\subsection{Performance Parameters}

To evaluate the performance of the GEM detector, three parameters are considered: the effective gain, ion backflow (IBF), and the ion backflow-to-gain (IBF/Gain) ratio.

\textbf{Effective Gain:}
The effective gain is defined as the ratio of the number of electrons collected at the readout electrode to the number of incident particles entering the detector.
\begin{equation}
G_{\mathrm{eff}}=\frac{N_{\mathrm{collected}}}{N_{\mathrm{incident}}}.
\end{equation}
The electron multiplication inside the GEM holes is governed by the Townsend ionization and can be approximated by

\begin{equation}
G \approx \exp \left(\int \alpha(E)\,dx\right),
\end{equation}

where $\alpha(E)$ is the Townsend ionization coefficient, which depends on the local electric field.

\textbf{Ion Backflow (IBF):}
Ion backflow is defined as the fraction of positive ions generated during avalanche multiplication that drift back into the drift region. It is calculated as

\begin{equation}
\mathrm{IBF}=\frac{N_{\mathrm{ions}}^{\mathrm{drift}}}{N_{\mathrm{incident}}}.
\end{equation}

Since each avalanche electron is accompanied by a positive ion, the ion backflow is closely related to the avalanche multiplication process and the electric-field configuration inside the detector.

\textbf{The IBF/Gain Ratio:}
The ion-backflow-to-gain (IBF/Gain) ratio is a normalized performance parameter that quantifies the balance between charge amplification and ion suppression. A lower value of this ratio indicates better suppression of ion backflow for a given detector gain.
\section{Experimental validation}

To establish the reliability of the computational framework employed in this work, the gain of a conventional single GEM detector is validated against published experimental measurements~\cite{bachmann1999charge}. The present work uses the same validated framework reported in recent studies~\cite{rana2026}.

The detector consists of a standard bi-conical GEM foil with a hole pitch of 140~$\mu$m, an inner hole diameter of 50~$\mu$m, an outer hole diameter of 70~$\mu$m, and a 50~$\mu$m thick Kapton foil coated with 5~$\mu$m copper on both sides. The drift and induction fields are fixed at 1.5~kV/cm and 5~kV/cm, respectively, while the voltage difference between the GEM foil ($\Delta V_{\mathrm{GEM}}$) is varied. The detector is irradiated with a 6~keV X-ray source, producing approximately 220 primary electrons in the gas.

The electrostatic field distribution is computed in the ANSYS Mechanical APDL using the same detector geometry and electric-field configuration. The computed field maps are imported into Garfield$^{++}$, where microscopic electron transport and avalanche multiplication are calculated under the same operating conditions. The effective gain is obtained by calculating the ratio of the total number of avalanche electrons to the number of primary electrons.

 \begin{figure}[H]
  \centering
  \includegraphics[width=0.9\columnwidth]{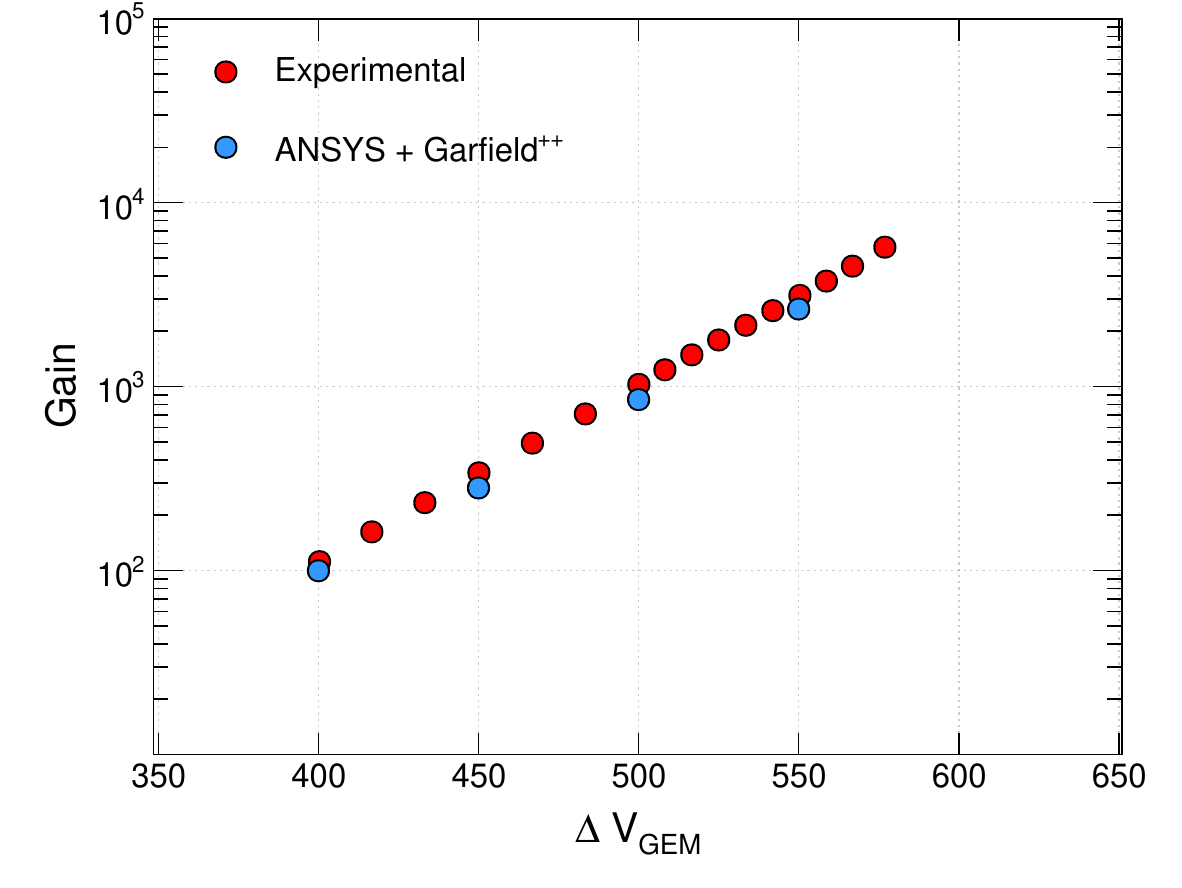}
  \caption{Effective gain of a conventional single GEM detector as a function of the voltage difference across the GEM foil ($\Delta V_{\mathrm{GEM}}$). The comparison between the experimental measurements (red circles) and the calculated results (blue circles) demonstrates good agreement, validating the framework.} 
  \label{fig:validation}
\end{figure}
The calculated effective gain agrees well with the experimental measurements for different values of $\Delta V_{\mathrm{GEM}}$, as shown in Fig.~\ref{fig:validation}. This agreement across the entire voltage range shows that the framework correctly reproduces the electron multiplication process in a standard GEM detector. The modest discrepancies between the calculated and experimental values are within the expected experimental and statistical uncertainties and do not influence the overall agreement between the two data sets.

The agreement between the calculated and experimental data confirms the validity of the ANSYS-Garfield$^{++}$ framework and demonstrates its applicability to the investigation of different GEM hole geometries and their response to various charged particles.

\section{Results and Discussion}
The response of a gaseous detector is fundamentally determined by the primary ionization produced by an incident charged particle. The mean energy loss per unit path length in the gas is described by the Bethe–Bloch equation:
\begin{equation}
-\frac{dE}{dx} = K z^2 \frac{Z}{A} \frac{1}{\beta^2} 
\left[ \frac{1}{2} \ln \left( \frac{2 m_e c^2 \beta^2 \gamma^2 W_{\mathrm{max}}}{I^2} \right) 
- \beta^2 - \frac{\delta}{2} \right],
\end{equation}
where z is the charge of the incident particle, $\beta$ and $\gamma$ are relativistic factors, Z and A are the atomic number and mass of the gas, I is the mean excitation energy, and W$_\mathrm{max}$ is the maximum energy transfer per collision.
The number of primary electron–ion pairs produced per unit length is given by:
\begin{equation}
N_{\mathrm{prim}} = \frac{1}{W} \frac{dE}{dx},
\end{equation}
where W is the mean energy required to produce an electron–ion pair in the gas medium. The detector response is governed by two competing processes: electron multiplication inside the GEM holes and the transport of positive ions produced during the avalanche. Although a higher gain improves the signal amplitude, excessive ion backflow can distort the electric field in the drift region and degrade detector performance during long-term operation. This study examines the dependence of effective gain and ion backflow on particle type and incident energy, and compares the detector response for conventional bi-conical and optimized single-conical GEM geometries.

The detector response to muons, pions, kaons, and protons was examined over an incident-energy range extending up to 10 GeV. These particles span a wide energy range, providing a suitable benchmark for evaluating the detector response under realistic conditions encountered in collider experiments.

\subsection{Single Gem characteristics}
\begin{figure}[H]
  \centering
  \includegraphics[width=\columnwidth]{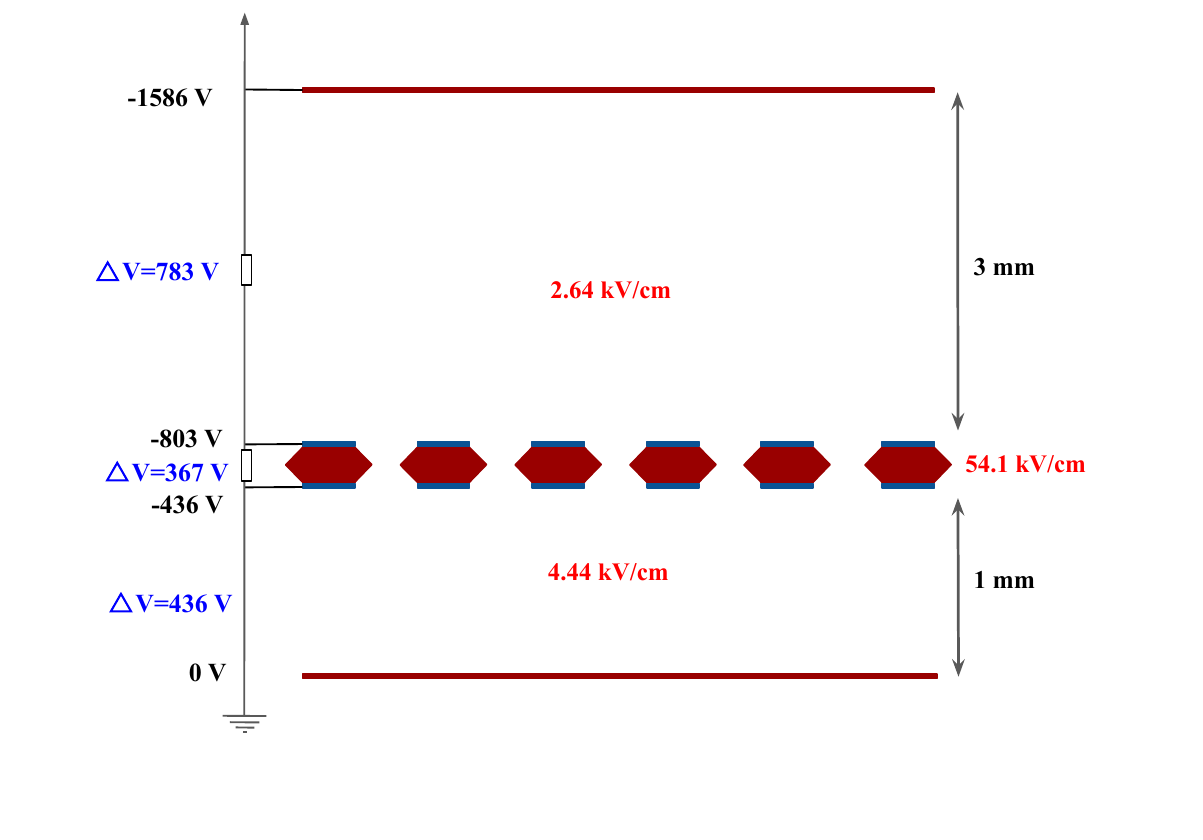}
  \caption{Operating configuration of the conventional bi-conical single-GEM detector illustrating the detector geometry, applied electrode voltages, voltage differences across the detector, and the resulting electric-field strengths in the drift, amplification, and induction regions.}
  \label{fig:SSG}
\end{figure}

The response characteristics of the single GEM detector were initially studied for both the standard bi-conical and optimized single-conical hole geometries to understand the basic behavior of charge amplification and ion transport within a single foil. The detector geometry with the applied voltages and the corresponding electric-field strengths in the drift, amplification and induction region is shown in Fig~\ref{fig:SSG}.

The drift field guides the primary electrons towards the GEM holes without initiating multiplication, while the induction field efficiently extracts the avalanche electrons towards the readout plane. An electric field of about 54.1 kV/cm is established inside the GEM holes, where electron multiplication takes place. The voltage configuration was kept fixed throughout the study so that the observed variations in the detector response arise only from differences in the incident particle and its energy. 
Since a single GEM isolates a single amplification stage, it provides a direct way to study the effects of particle-dependent ionization and hole geometry on detector performance.
%The simulations were performed for muons, pions, kaons, and protons over a range of incident energies upto 10 GeV.

  \begin{figure}[H]
  \centering
  \includegraphics[width=\columnwidth]{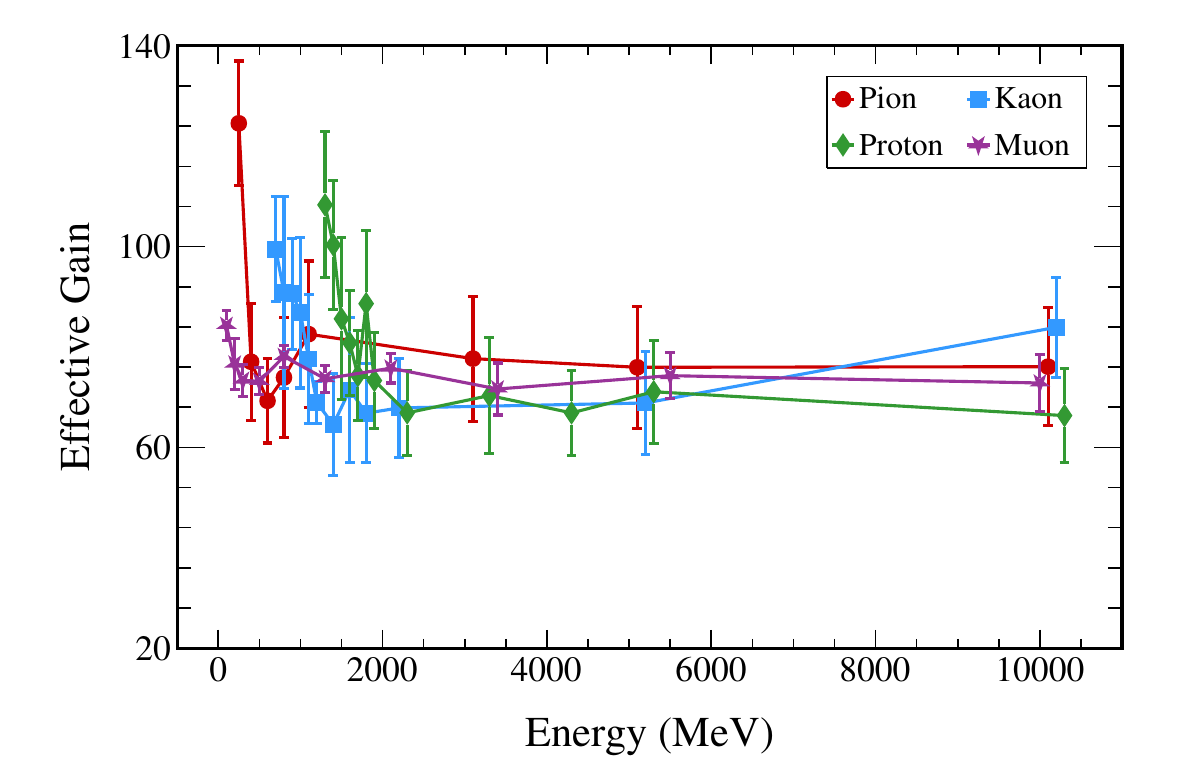}
  \caption{Variation of the effective gain with incident particle energy for muons, pions, kaons, and protons in the conventional bi-conical single-foil GEM detector. The points represent the calculated effective gain, while the lines are included to guide the eye. Error bars represent the statistical uncertainties obtained from the Garfield$^{++}$ calculation.}
  \label{fig:SSG_gain}
\end{figure}

The effective gain obtained for the standard single GEM is shown in Fig.~\ref{fig:SSG_gain}. The error bars shown in all figures throughout this study represent the statistical uncertainties obtained from Garfield$^{++}$. The gain exhibits a weak dependence on the particle species and incident energy throughout the investigated range. Although protons, kaons, pions, and muons produce different numbers of primary ionization electrons because of their different stopping powers, the multiplication inside the GEM hole is controlled mainly by the local electric field. Once the primary electrons enter the high-field region, they undergo nearly identical avalanche multiplication regardless of the particle that produced them. Consequently, the effective gain remains almost unchanged for all particle species. A slight difference is observed at lower energies, where the variation in stopping power is predicted by the Bethe-Bloch equation. As the particle energy increases, the stopping powers gradually converge towards the minimum-ionizing region, resulting in a more uniform detector response.

  \begin{figure}[!t]
  \centering
  \includegraphics[width=\columnwidth]{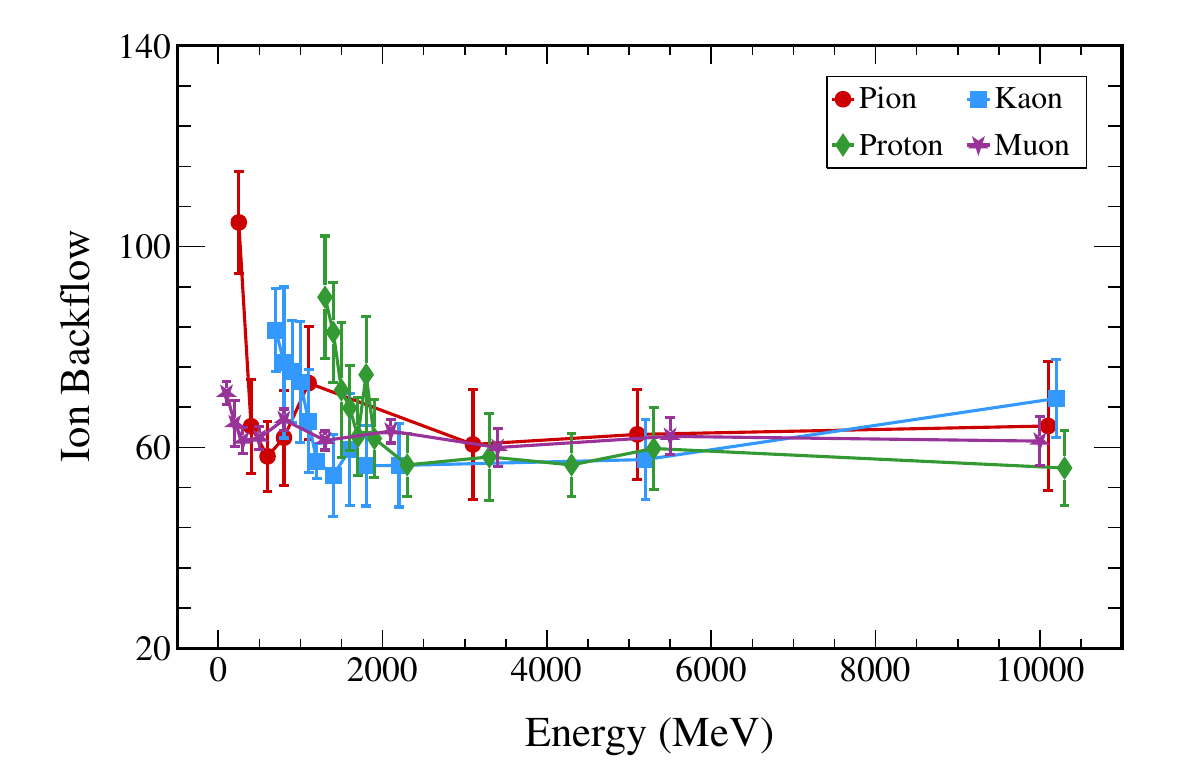}
  \caption{Ion-backflow as a function of the incident particle energy for muons, pions, kaons, and protons in the standard bi-conical single-foil GEM detector.}
  \label{fig:SSG_ib}
\end{figure}

The corresponding ion-backflow is shown in Fig.~\ref{fig:SSG_ib} and follows the same trend as the effective gain, since each avalanche electron is accompanied by a positive ion. Most ions are collected by the GEM electrodes, with only a small fraction drifting back into the drift region, whereas the small differences among particle species arise mainly from variations in primary ionization. 
  \begin{figure}[H]
  \centering
  \includegraphics[width=\columnwidth]{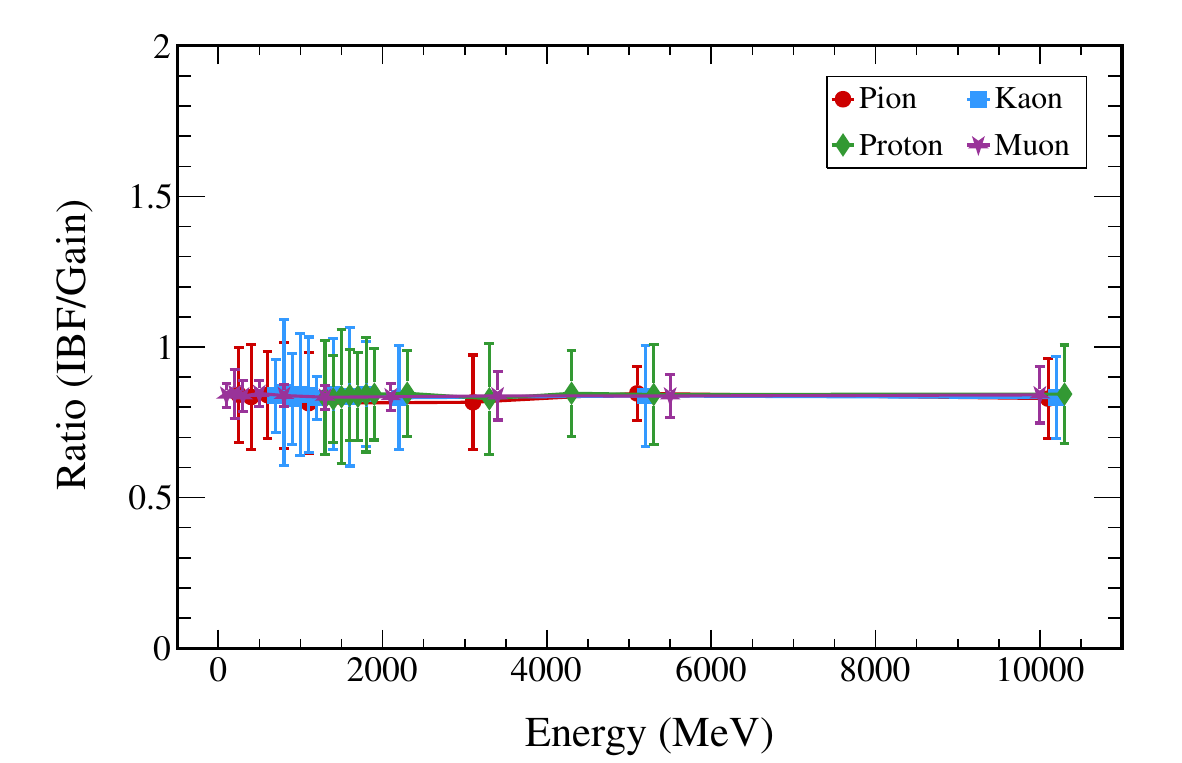}
  \caption{Ion-backflow/gain ratio as a function of incident particle energy for muons, pions, kaons, and protons in the standard bi-conical single-foil GEM detector.}
  \label{fig:SSG_ratio}
\end{figure}

To evaluate the trade-off between charge amplification and ion suppression, the IBF/Gain ratio (Fig.~\ref{fig:SSG_ratio}) is used as a normalized performance metric. %The ratio remains nearly constant across different particle species and incident energies, indicating that electron multiplication and ion backflow vary in nearly the same proportion and are governed primarily by the electric-field configuration and hole geometry.
The ratio remains nearly constant across different particles and incident energies, indicating that electron multiplication and ion backflow increase in nearly equal proportions and are determined primarily by the electric-field configuration and hole geometry. Table~\ref{tab:single_standard_muon} shows the  numerical values of the effective gain, ion backflow, and IBF/Gain ratio for the muon at different energies, which serves as a baseline for comparison with the optimized GEM geometry.

\begin{table}[h]
\centering
\renewcommand{\arraystretch}{1.4}
\begin{tabular}{|c|c|c|c|}
\hline
\textbf{Energy(MeV)} & \textbf{Gain} & \textbf{Ion Backflow} & \textbf{Ratio} \\
\hline
100   & $84.29 \pm 2.97$ & $70.76 \pm 2.29$ & $0.840 \pm 0.040$ \\
\hline
300   & $73.36 \pm 3.15$ & $61.41 \pm 2.57$ & $0.837 \pm 0.050$ \\
\hline
800   & $78.11 \pm 2.11$ & $65.52 \pm 2.12$ & $0.839 \pm 0.035$ \\
\hline
3400  & $71.63 \pm 5.19$ & $59.97 \pm 3.76$ & $0.838 \pm 0.080$ \\
\hline
10000 & $72.83 \pm 5.70$ & $61.25 \pm 4.83$ & $0.841 \pm 0.093$ \\
\hline
\end{tabular}
\renewcommand{\arraystretch}{1}
\caption{Effective gain, ion backflow, and IBF/Gain ratio of the standard single-GEM configuration for muons.}
\label{tab:single_standard_muon}
\end{table}

  \begin{figure}[!t]
  \centering
  \includegraphics[width=\columnwidth]{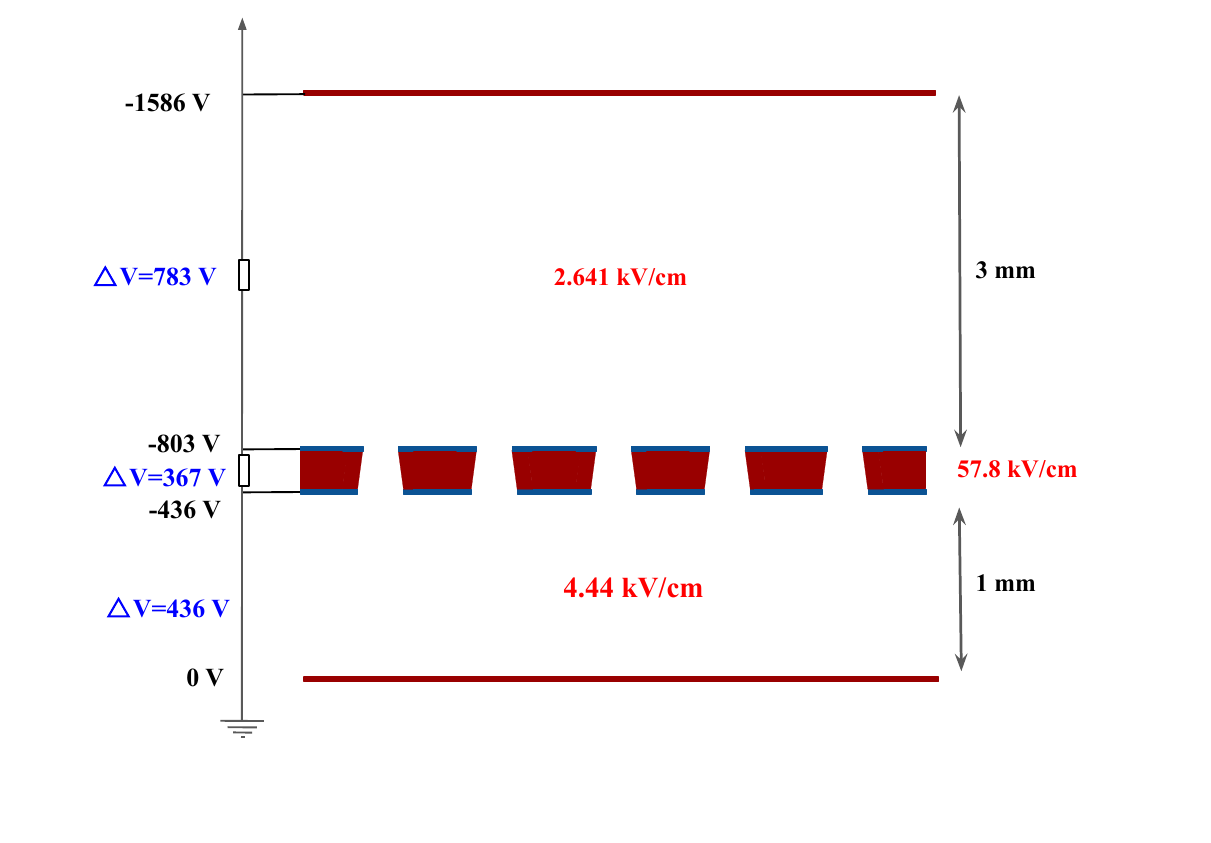}
  \caption{Operating configuration of the optimized single-conical single-GEM detector illustrating the detector geometry, applied electrode voltages, voltage differences across the detector, and the resulting electric-field strengths in the drift, amplification, and induction regions.}
  \label{fig:MSG}
\end{figure}
\begin{figure}[!t]
  \centering
  \includegraphics[width=\columnwidth]{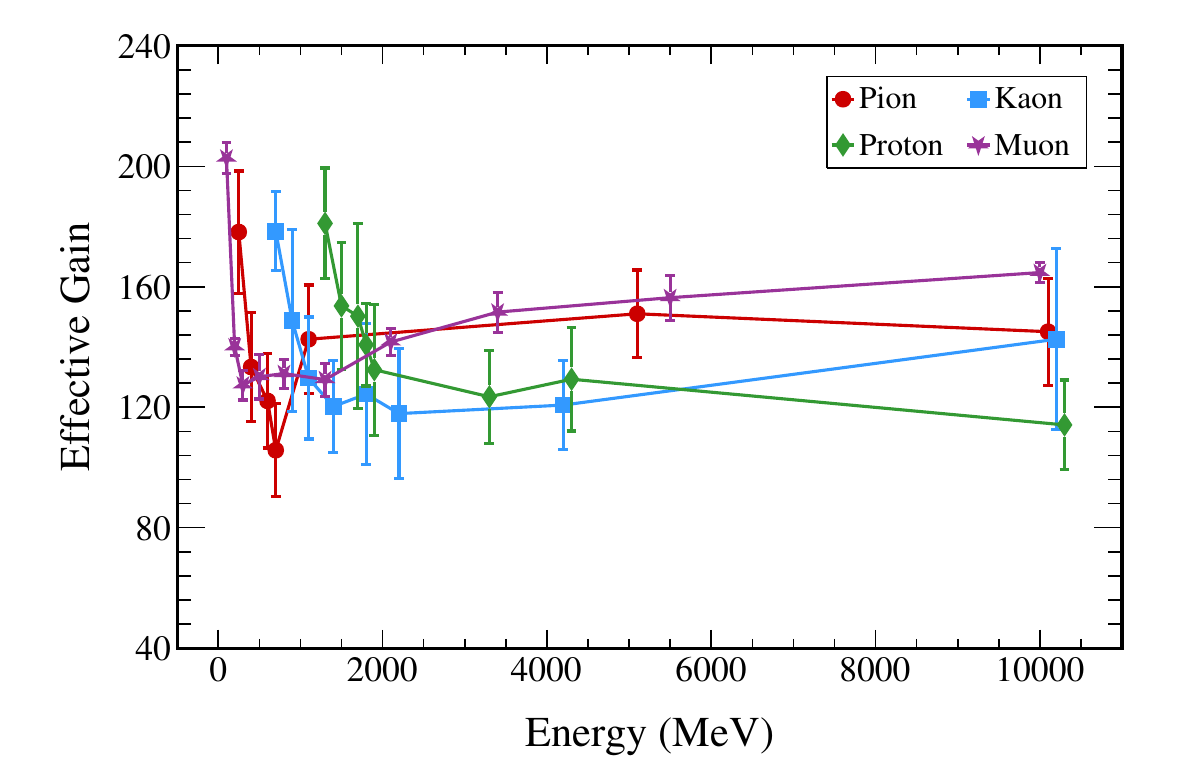}
  \caption{Effective gain as a function of incident particle energy for muons, pions, kaons, and protons in the optimized single-conical single-foil GEM detector.}
  \label{fig:SCSG_gain}
\end{figure}
\begin{figure}[!h]
  \centering
  \includegraphics[width=\columnwidth]{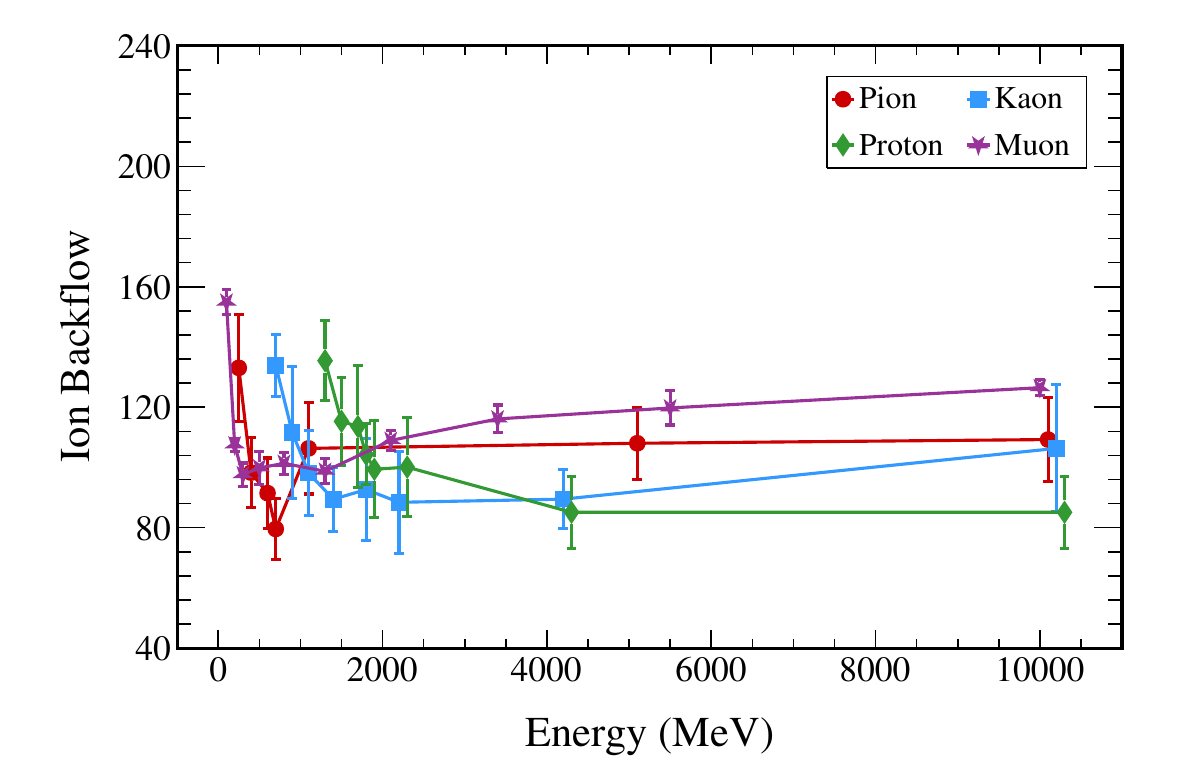}
  \caption{Ion-backflow as a function of incident particle energy for muons, pions, kaons, and protons in the optimized single-conical single-foil GEM detector.}
  \label{fig:SCSG_ib}
\end{figure}

To investigate the influence of hole geometry on detector performance, the bi-conical GEM was replaced with a single-conical design while keeping the detector dimensions, gas mixture, and operating voltages unchanged. Fig.~\ref{fig:MSG} shows the optimized single-GEM with the corresponding electric-field distribution. Compared with the conventional bi-conical geometry, the electric field inside the single-conical hole increases from about 54.1 kV/cm to 57.8 kV/cm, while the drift and induction fields remain the same. A higher electric field leads to stronger electron multiplication, resulting in increased gain. This indicates that the improvement originates from optimized hole geometry rather than the particle-dependent ionization process. In the modified hole geometry, the upper copper electrode is closer to the avalanche region, allowing positive ions to reach the electrode over a shorter distance. As a result, more ions are collected on the electrode, reducing ion backflow. Fig~\ref{fig:SCSG_gain} and~\ref{fig:SCSG_ib} demonstrate the variation of effective gain and ion-backflow, respectively, with incident energy for the optimized single conical geometry.

Suppression of ion backflow is particularly important for GEM detectors operating at high particle rates. The buildup of positive ions in the drift region produces space-charge effects that distort the electric field~\cite{Boehmer2013}, reduce electron transport efficiency, and degrade detector stability and tracking performance. In addition, the asymmetric hole profile guides the avalanche ions more efficiently towards the GEM electrodes, reducing the ion accumulation within the holes and mitigating local distortions of the electric field.

\begin{figure}[!t]
  \centering
  \includegraphics[width=\columnwidth]{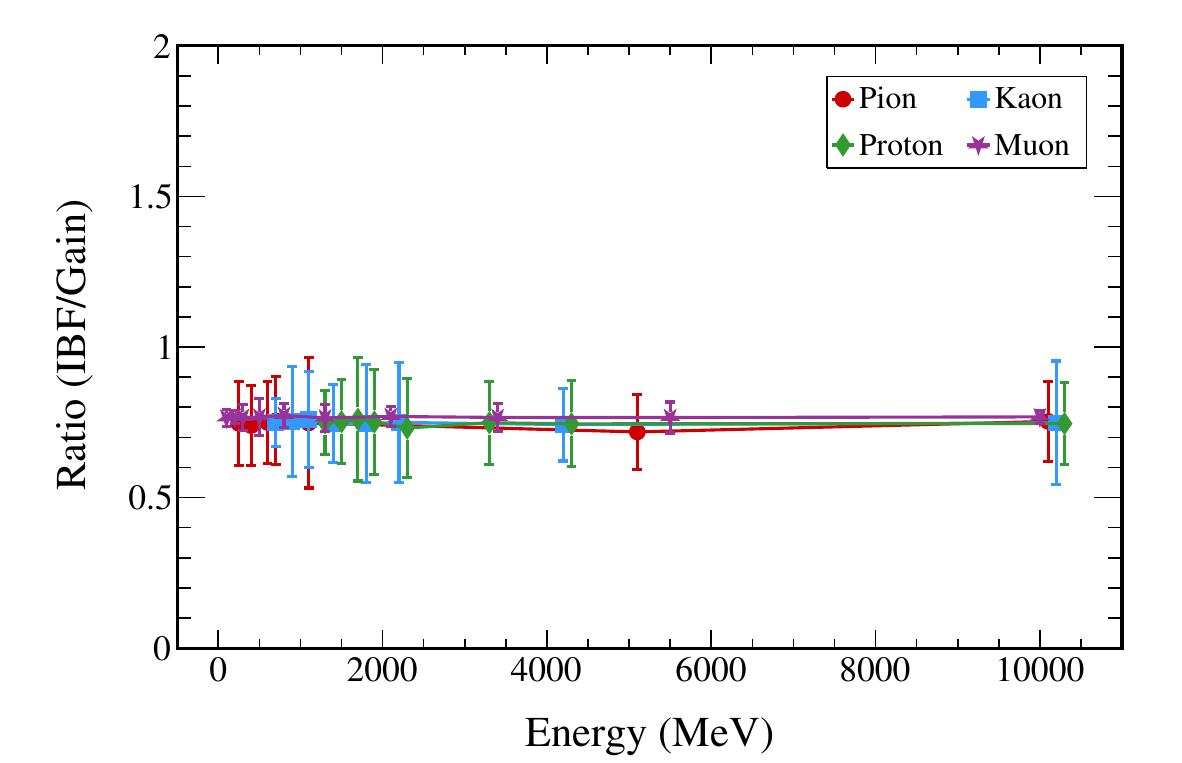}
  \caption{Ion-backflow/gain ratio as a function of incident particle energy for muons, pions, kaons, and protons in the optimized single-conical single-foil GEM detector.}
  \label{fig:MSG_ratio}
\end{figure}

The IBF/Gain ratio shown in Fig.~\ref{fig:MSG_ratio} gives a clearer measure of detector performance. Compared with conventional single-GEM geometry, the optimized single-conical geometry exhibits an increase in ion backflow, accompanied by a substantially larger increase in effective gain over a wide energy range. As a result, the IBF/Gain ratio is consistently lower for the optimized geometry, indicating that the increase in gain outweighs the corresponding increase in ion backflow. This reduction in the IBF/Gain ratio is observed for all particle species studied. Table~\ref{tab:single_modified_muon} gives the summary of effective gain, ion backflow, and IBF/Gain ratio obtained for muons at different incident energies in the optimized single-conical single-foil GEM detector. Similar trends are observed for pions, kaons, and protons over the studied energy range. These observations motivated the implementation of the proposed geometry in a triple-GEM detector to evaluate its performance under realistic operating conditions.

\begin{table}[h]
\centering
\small
\renewcommand{\arraystretch}{1.3}
\begin{tabular}{|c|c|c|c|}
\hline
\textbf{Energy(MeV)} & \textbf{Gain} & \textbf{Ion Backflow} & \textbf{Ratio} \\
\hline
100   & $202.71 \pm 5.08$ & $154.88 \pm 4.19$ & $0.764 \pm 0.028$ \\
\hline
300   & $127.37 \pm 4.95$ & $97.59 \pm 3.88$  & $0.766 \pm 0.043$ \\
\hline
800   & $131.17 \pm 4.81$ & $101.35 \pm 3.63$ & $0.773 \pm 0.040$ \\
\hline
3400  & $151.59 \pm 6.65$ & $116.11 \pm 4.62$ & $0.766 \pm 0.045$ \\
\hline
10000 & $164.72 \pm 3.36$ & $126.49 \pm 2.59$ & $0.768 \pm 0.022$ \\
\hline
\end{tabular}
\renewcommand{\arraystretch}{1}
\caption{Effective gain, ion backflow, and IBF/Gain ratio of the optimized single-GEM configuration for muons.}
\label{tab:single_modified_muon}
\end{table}

\subsubsection{Triple GEM characteristics}
Although the single GEM configuration provides insight into the microscopic amplification behavior of an individual foil, modern high-energy physics detectors generally employ multiple GEM stages to achieve high, stable gain while minimizing ion backflow. Fig.~\ref{fig:STG} illustrates the standard triple-GEM detector together with the applied voltages and the strength of the electric field in the drift, transfer, and induction regions. The electric fields were optimized to ensure efficient electron collection into the first GEM, high electron transparency between successive GEM stages, and effective extraction of avalanche electrons towards the readout plane.

\begin{figure}[!t]
  \centering
  \includegraphics[width=\columnwidth]{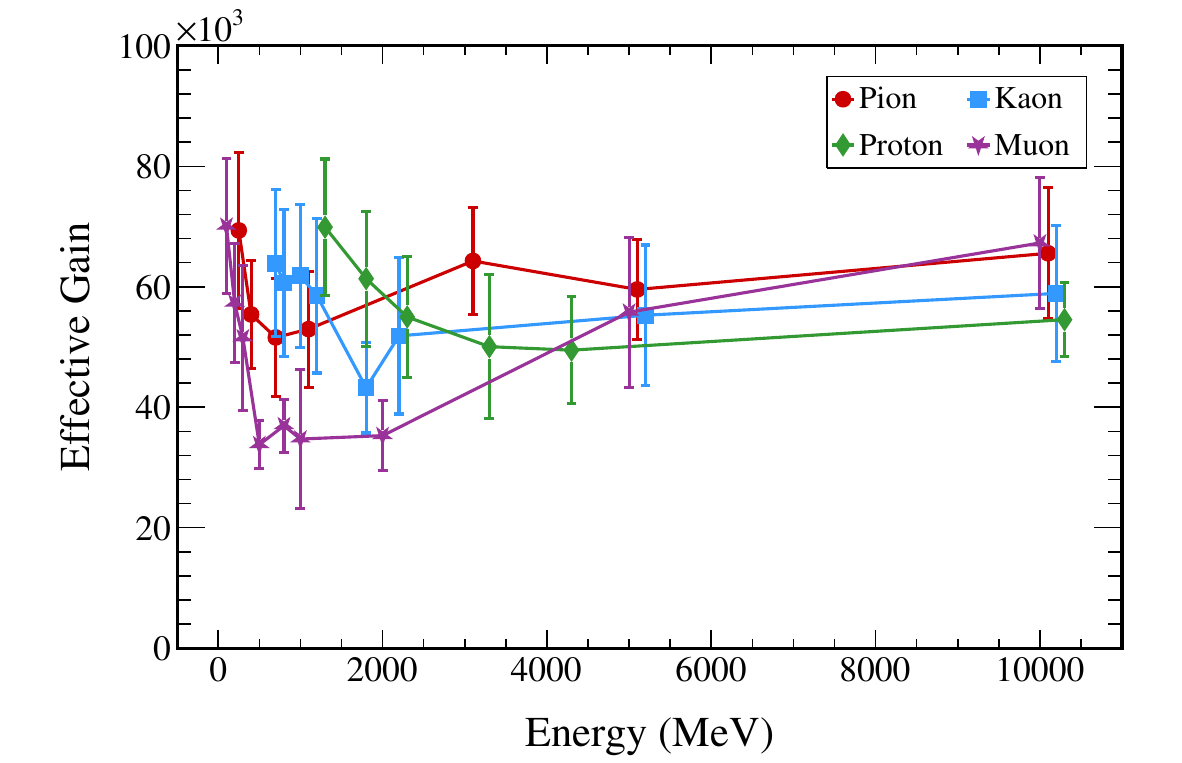}
  \caption{Effective gain as a function of incident particle energy for muons, pions, kaons, and protons in the standard bi-conical triple-foil GEM detector.}
  \label{fig:BCTG_gain}
\end{figure}

\begin{figure}[H]
  \centering
  \includegraphics[width=\columnwidth]{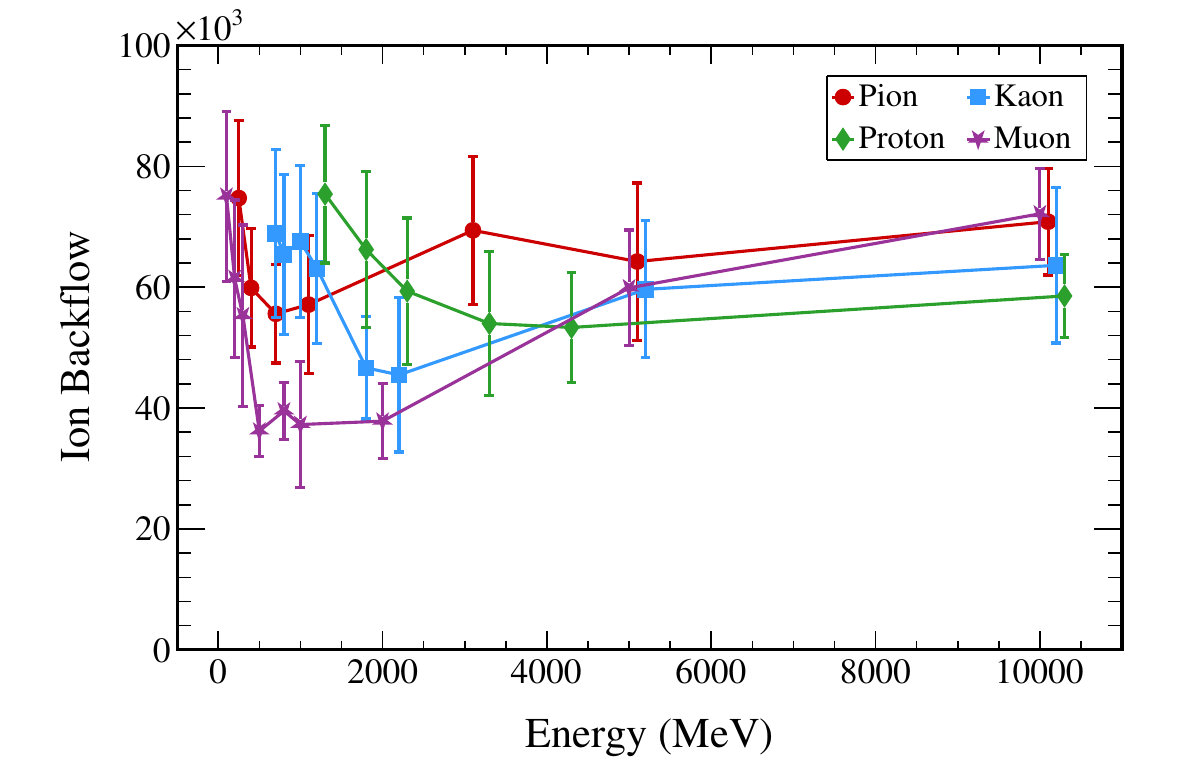}
  \caption{Ion-backflow as a function of incident particle energy for muons, pions, kaons, and protons in the standard bi-conical triple-foil GEM detector.}
  \label{fig:BCTG_ib}
\end{figure}

The effective gain and the corresponding ion backflow for the standard triple-GEM detector are presented in Fig.~\ref{fig:BCTG_gain} and ~\ref{fig:BCTG_ib}, respectively. As expected, cascading three GEM foils increases the overall gain by several orders of magnitude compared with the single-GEM configuration while retaining the weak dependence on particle species and incident energy. Because each avalanche electron is accompanied by a positive ion, the ion backflow follows a trend similar to the gain. Despite the significantly larger avalanche, the multi-stage configuration efficiently collects most of the ions on the intermediate GEM electrodes, allowing only a small fraction to drift back into the drift region. The small variations observed among different particle species originate primarily from differences in their initial ionization, whereas avalanche multiplication and ion transport remain governed mainly by the electric-field configuration within the GEM holes.

\begin{figure}[!t]
  \centering
  \includegraphics[width=\columnwidth]{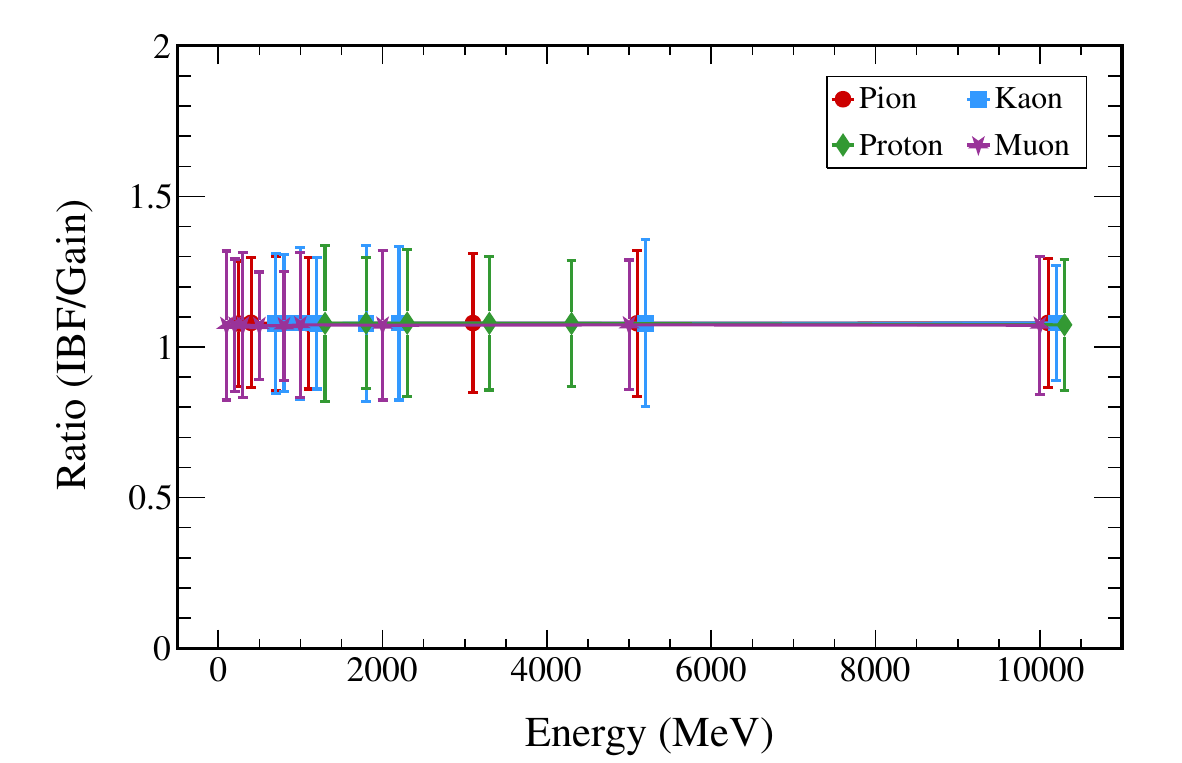}
  \caption{Ion-backflow/gain ratio as a function of incident particle energy for muons, pions, kaons, and protons in the standard bi-conical triple-foil GEM detector.}
  \label{fig:BCTG_ratio}
\end{figure}

The IBF/Gain ratio shown in Fig.~\ref{fig:BCTG_ratio} remains nearly constant throughout the energy for all particle species. The numerical values of the effective gain, ion backflow, and IBF/Gain ratio for the muon response at selected energies are summarized in Table~\ref{tab:triple_standard_muon}.

\begin{table}[h]
\centering
\renewcommand{\arraystretch}{1.4}
\begin{tabular}{|c|c|c|c|}
\hline
\textbf{\shortstack{Energy \\ (MeV)}} & \textbf{\shortstack{Gain \\ $\boldsymbol{(\times 10^3)}$}} & \textbf{\shortstack{Ion Backflow \\ $\boldsymbol{(\times 10^3)}$}} & \textbf{\shortstack{Ratio \\ \vphantom{Ion Backflow}}} \\
\hline
100   & $70.04 \pm 11.21$ & $75.05 \pm 14.04$ & $1.071 \pm 0.248$ \\
\hline
300   & $51.52 \pm 12.00$ & $55.27 \pm 15.03$ & $1.073 \pm 0.241$ \\
\hline
1000  & $34.72 \pm 11.52$ & $37.27 \pm 10.39$ & $1.074 \pm 0.240$ \\
\hline
2000  & $35.29 \pm 5.78$  & $37.84 \pm 6.19$  & $1.072 \pm 0.248$ \\
\hline
10000 & $67.27 \pm 10.90$ & $72.14 \pm 8.93$  & $1.072 \pm 0.229$ \\
\hline
\end{tabular}
\renewcommand{\arraystretch}{1}
\caption{Effective gain, ion backflow, and IBF/Gain ratio, with associated uncertainties, of the Standard triple-GEM configuration for muons across varying energies.}
\label{tab:triple_standard_muon}
\end{table}

   \begin{figure}[t!]
  \centering
  \includegraphics[width=\columnwidth]{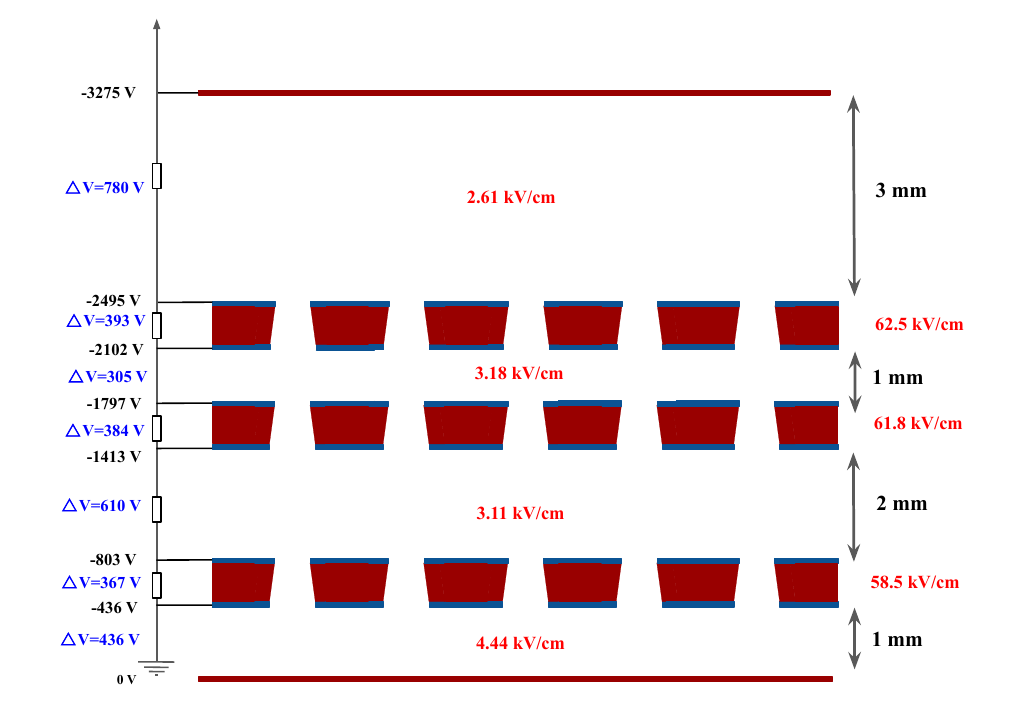}
  \caption{Operating configuration of the optimized single-conical triple-GEM detector illustrating the detector geometry, applied electrode voltages, voltage differences across the detector, and the resulting electric-field strengths in the drift, amplification, and induction regions.}
  \label{fig:MTG}
\end{figure}

 Finally, the proposed single-conical hole geometry was implemented in all three GEM foils while maintaining the same detector dimensions, gas composition, and operating voltages as the conventional triple-GEM detector.

 Fig.~\ref{fig:MTG} shows the corresponding detector configuration and electric-field distribution. The modified geometry preserves the electric-field structure required for stable detector operation while altering the field distribution inside each GEM hole, thereby influencing both electron and ion transport throughout the multi-stage amplification process.
 
  \begin{figure}[!t]
  \centering
  \includegraphics[width=\columnwidth]{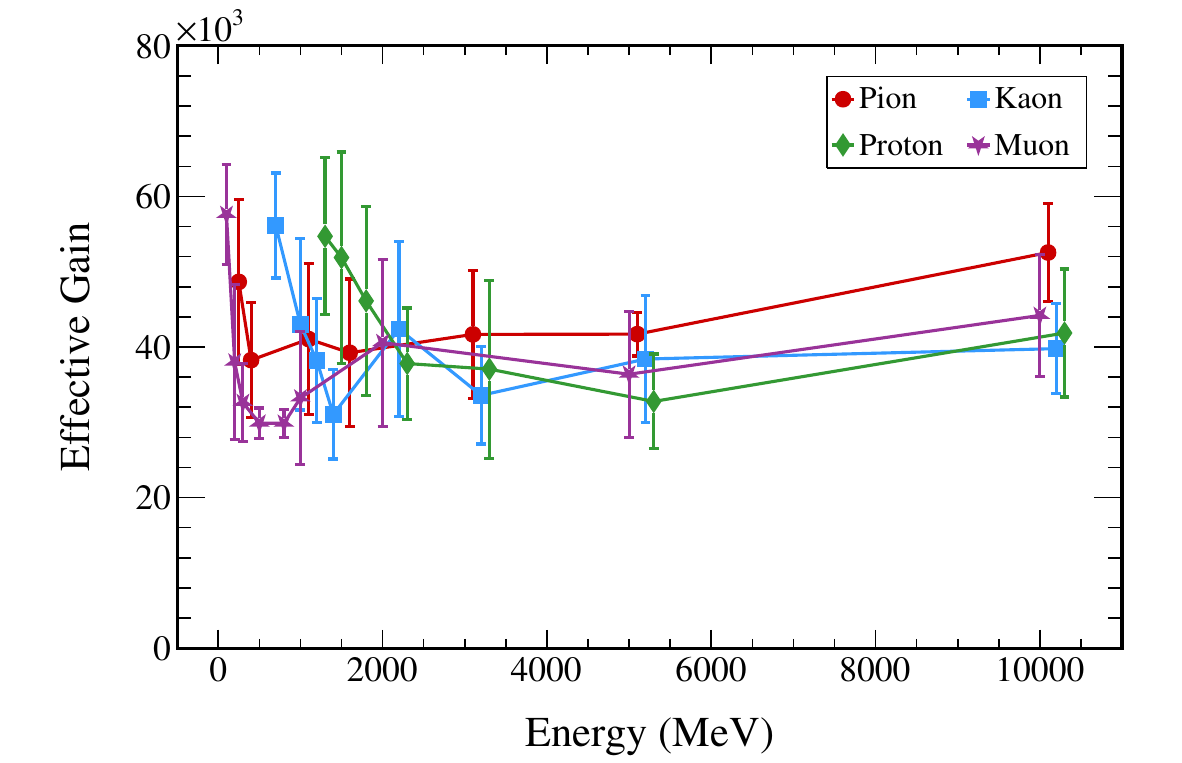}
  \caption{Effective gain as a function of incident particle energy for muons, pions, kaons, and protons in the optimized single-conical triple-foil GEM detector.}
  \label{fig:SCTG_gain}
\end{figure}

\begin{figure}[!h]
  \centering
  \includegraphics[width=\columnwidth]{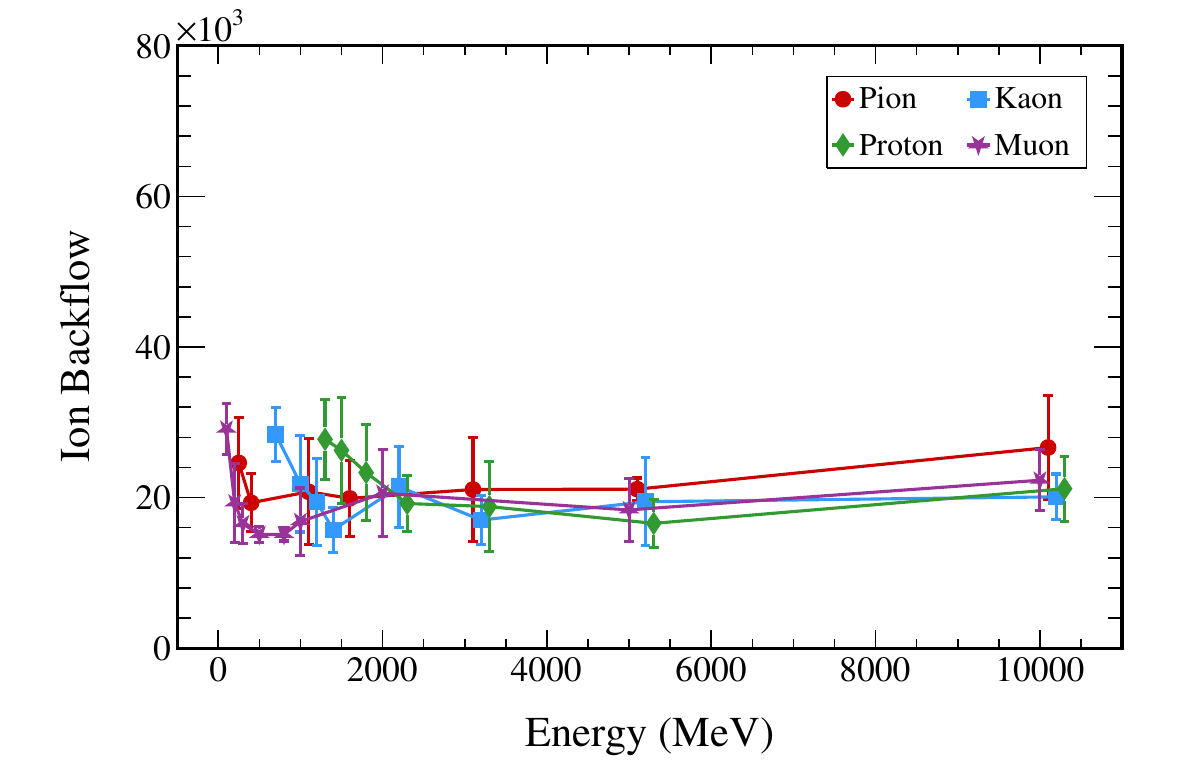}
  \caption{Ion-backflow as a function of incident particle energy for muons, pions, kaons, and protons in the optimized single-conical triple-foil GEM detector.}
  \label{fig:SCTG_ib}
\end{figure}

The effective gain and the corresponding ion backflow obtained with the optimized triple-GEM detector are presented in Fig.~\ref{fig:SCTG_gain} and ~\ref{fig:SCTG_ib}, respectively. The reduced electron transparency associated with the smaller entrance diameter limits charge transfer between successive GEM foils, producing a modest reduction in effective gain despite the stronger local electric field. A similar effect is observed for all the particle species. %This reduction is attributed to the lower electron transparency of the selected single-conical hole geometry, where a smaller fraction of electrons is efficiently transferred between successive GEM stages. 
By appropriately adjusting the upper and lower hole diameters to maintain electron transparency comparable to that of the standard GEM, the gain can be further improved.
Despite the modest decrease in gain, the modified geometry exhibits significantly lower ion backflow. The asymmetric electric-field distribution promotes more efficient collection of avalanche ions on the GEM electrodes, reducing the number of ions drifting back into the drift region.

  \begin{figure}[!t]
  \centering
  \includegraphics[width=\columnwidth]{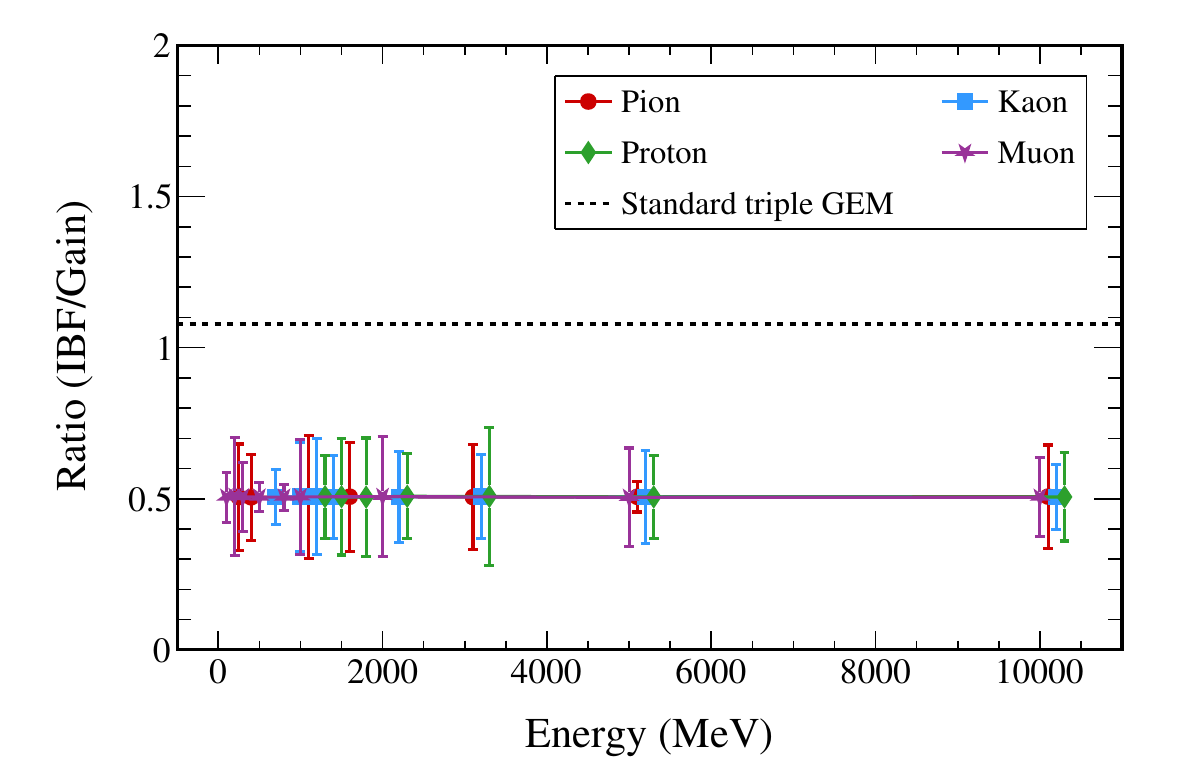}
  \caption{Ion-backflow/gain ratio as a function of incident particle energy for muons, pions, kaons, and protons in the optimized single-conical triple-foil GEM detector.}
  \label{fig:MTG_plot}
\end{figure}

The improvement becomes more evident from the IBF/Gain ratio shown in Fig.~\ref{fig:MTG_plot}. Table~\ref{tab:triple_modified_muon} summarizes the corresponding effective gain, ion backflow, and IBF/Gain ratio for the optimized single-conical triple-GEM detectors for incident muons at different energies. The gain decreases slightly due to reduced electron transparency in the modified geometry.

\begin{table}[!h]
\centering
\renewcommand{\arraystretch}{1.4}
\begin{tabular}{|c|c|c|c|}
\hline
\textbf{\shortstack{Energy \\ (MeV)}} & \textbf{\shortstack{Gain \\ $\boldsymbol{(\times 10^3)}$}} & \textbf{\shortstack{Ion Backflow \\ $\boldsymbol{(\times 10^3)}$}} & \textbf{\shortstack{Ratio \\ \vphantom{Ion Backflow}}} \\
\hline
100   & $57.60 \pm 6.64$  & $29.08 \pm 3.38$ & $0.505 \pm 0.083$ \\
\hline
300   & $32.62 \pm 5.19$  & $16.52 \pm 2.64$ & $0.506 \pm 0.114$ \\
\hline
1000  & $33.22 \pm 8.82$  & $16.81 \pm 4.47$ & $0.506 \pm 0.190$ \\
\hline
2000  & $40.58 \pm 11.09$ & $20.58 \pm 5.77$ & $0.507 \pm 0.199$ \\
\hline
10000 & $44.19 \pm 8.15$  & $22.30 \pm 4.05$ & $0.505 \pm 0.131$ \\
\hline
\end{tabular}
\renewcommand{\arraystretch}{1}
\caption{Effective gain, ion backflow, and IBF/Gain ratio, with associated uncertainties, of the Modified triple-GEM configuration for muons across five energies.}
\label{tab:triple_modified_muon}
\end{table}

This reduction is reasonably uniform with an average improvement of approximately 52\% over the standard bi-conical design shown in Table~\ref{tab:triple_ratio_improvement}. The ratio is almost independent of the particle type, indicating that the improvement is due to the changed hole geometry. This demonstrates that the optimized single-conical geometry gives a better balance between charge amplification and ion suppression than the conventional bi-conical design.

\begin{table}[h]
\centering
\renewcommand{\arraystretch}{1.3}
\begin{tabular}{|c|c|c|c|}
\hline
\textbf{\shortstack{Energy \\ (MeV)}} & \textbf{\shortstack{Ratio \\ (Standard)}} & \textbf{\shortstack{Ratio \\ (Optimised)}} & \textbf{\shortstack{Improvement \\ (in \%)}} \\\hline
100   & $1.071 \pm 0.248$ & $0.505 \pm 0.083$ & 52.88\% \\
\hline
300   & $1.073 \pm 0.241$ & $0.506 \pm 0.114$ & 52.79\% \\
\hline
1000  & $1.074 \pm 0.240$ & $0.506 \pm 0.190$ & 52.88\% \\
\hline
2000  & $1.072 \pm 0.248$ & $0.507 \pm 0.199$ & 52.71\% \\
\hline
10000 & $1.072 \pm 0.229$ & $0.505 \pm 0.131$ & 52.94\% \\
\hline
\end{tabular}
\renewcommand{\arraystretch}{1}
\caption{Percentage improvement in the ion-backflow-to-gain ratio between the Standard and Optimized triple-GEM configurations for muons. Average improvement: 52.84$\%$}
\label{tab:triple_ratio_improvement}
\end{table}

\section{Conclusion}

A comprehensive investigation of the response of conventional and modified GEM detectors to muons, pions, kaons, and protons has been carried out using ANSYS and Garfield$^{++}$. The influence of particle-dependent ionization on effective gain and ion backflow was studied for both single and triple-GEM configurations over a wide energy range.

For the conventional geometries, both gain and ion backflow showed only a weak dependence on particle species and incident energy, indicating that the detector response is governed primarily by the electric-field distribution within the GEM holes. The proposed single-conical hole geometry modifies the field distribution and alters the transport of electrons and avalanche ions without changing the detector's operating conditions.

In the single-GEM configuration, the optimized geometry produced a slightly lower IBF/Gain ratio with a large increase in gain. When extended to a triple-GEM detector, the modified geometry resulted in a slight reduction in gain due to lower electron transparency between successive GEM stages. A higher gain can be achieved by increasing the upper-hole size while maintaining the same upper-to-lower-hole size ratio, provided that the electric-field configuration remains unchanged. However, this reduction was accompanied by a much larger decrease in ion backflow, leading to an average improvement of about 52\% in the ion backflow-to-gain ratio compared to the conventional bi-conical design.

These results demonstrate that optimization of the GEM hole geometry provides a significantly improved balance between charge amplification and ion suppression without requiring changes to the detector architecture or operating conditions. The proposed single-conical geometry enhances ion collection efficiency while preserving effective charge multiplication, thereby reducing the likelihood of space-charge accumulation under high-rate conditions. Because of its simple implementation, compatibility with single-mask fabrication, and improved performance, the proposed geometry represents a practical and promising design for next-generation GEM-based gaseous tracking detectors in future high-energy physics experiments.

%These results demonstrate that optimizing the GEM hole geometry provides a more favorable balance between charge amplification and ion suppression, reducing the likelihood of space-charge accumulation while preserving stable detector operation. The proposed single-conical geometry, therefore, represents a promising approach to improving the performance of GEM-based detectors operating in high-rate environments and may contribute to the development of future gaseous tracking detectors for high-energy physics experiments. Overall, the proposed single-conical geometry provides a noticeably improved balance between charge amplification and ion suppression without modifying the detector architecture or operating conditions. Such geometry optimization offers a practical route toward developing next-generation high-rate gaseous detectors for future collider experiments.

%These results demonstrate that microscopic optimization of the GEM hole geometry provides an effective route toward next-generation high-rate gaseous detectors without requiring changes to detector architecture or operating conditions.

\section*{Acknowledgment}

The authors appreciate the financial support provided by the Anusandhan National Research Foundation (ANRF), Government of India, under the Individual Research Grant (IRG) project \textbf{ANRF/IRG/2025/000338/PS}. The authors also acknowledge the Indian Institute of Technology (IIT) Mandi for providing the computational and research facilities necessary to carry out this work. The authors sincerely thank the members of the CMS and DRD1 Collaborations for their valuable discussions, insightful suggestions, and constructive feedback, which have greatly benefited this work.

\balance
\bibliographystyle{IEEEtran}
\bibliography{citations}

@IEEEtranBSTCTL{BSTcontrol,
  CTLdash_repeated_names = "no",
}

@article{Sauli1997GEM,
  author = {F. Sauli},
  title = {GEM: A New Concept for Electron Amplification in Gas Detectors},
  journal = {Nucl. Instrum. Meth. A},
  volume = {386},
  pages = {531--534},
  year = {1997},
  doi = {10.1016/S0168-9002(96)01172-2}
}

@article{Sauli2016Review,
  author = {F. Sauli},
  title = {The Gas Electron Multiplier (GEM): Operating Principles and Applications},
  journal = {Nucl. Instrum. Meth. A},
  volume = {805},
  pages = {2--24},
  year = {2016},
  doi = {10.1016/j.nima.2015.07.060}
}

@article{MPGDReview1999,
  author = {F. Sauli and A. Sharma},
  title = {Micro-pattern gaseous detectors},
  journal = {Annual Review of Nuclear and Particle Science},
  volume = {49},
  pages = {341--388},
  year = {1999},
  doi = {10.1146/annurev.nucl.49.1.341}
}

@article{Bouclier1997IEEE,
  author = {R. Bouclier et al.},
  title = {The Gas Electron Multiplier (GEM)},
  journal = {IEEE Trans. Nucl. Sci.},
  volume = {44},
  number = {3},
  pages = {646--650},
  year = {1997},
  doi = {10.1109/23.603726}
}

@article{MSGCProblems,
  author = {Fabio Sauli},
  title = {Principles of Operation of Multiwire Proportional and Drift Chambers},
  journal = {CERN Yellow Report},
  year = {1977}
}

@article{GEMApplications2003,
  author = {Fabio Sauli},
  title = {Development and Applications of Gas Electron Multiplier Detectors},
  journal = {Nucl. Instrum. Meth. A},
  volume = {505},
  pages = {195--198},
  year = {2003},
  doi = {10.1016/S0168-9002(03)01050-7}
}

@article{Mondal_2025,
doi = {10.1088/1748-0221/20/04/T04011},
year = {2025},
month = {apr},
publisher = {IOP Publishing},
volume = {20},
number = {04},
pages = {T04011},
author = {Mondal, Md Kaosor Ali and Angiras, Poojan and Rana, Sachin and Sarkar, A.},
title = {Optimizing Gas Electron Multiplier (GEM) geometry for enhanced gain, reduced ion backflow, and improved detector performance},
journal = {Journal of Instrumentation}
}

@article{GarfieldGeneral,
  author = {R. Veenhof},
  title = {Garfield, recent developments},
  journal = {Nucl. Instrum. Meth. A},
  year = {1998}
}

@article{PDGBetheBloch,
  author = {Particle Data Group},
  title = {Passage of Particles Through Matter},
  journal = {Prog. Theor. Exp. Phys.},
  year = {2020}
}

@manual{ANSYS2023,
  author       = {{ANSYS Inc.}},
  title        = {{ANSYS Mechanical APDL Documentation}},
  organization = {ANSYS Inc.},
  address      = {Canonsburg, PA, USA},
  year         = {2023},
  note         = {Release 2023 R1}
}

@article{CHARPAK200226,
title = {Micromegas, a multipurpose gaseous detector},
journal = {Nuclear Instruments and Methods in Physics Research Section A: Accelerators, Spectrometers, Detectors and Associated Equipment},
volume = {478},
number = {1},
pages = {26-36},
year = {2002},
note = {Proceedings of the ninth Int.Conf. on Instrumentation},
issn = {0168-9002},
doi = {https://doi.org/10.1016/S0168-9002(01)01713-2},
author = {G Charpak and J Derré and Y Giomataris and Ph Rebourgeard}
}

@article{PATRA201725,
title = {Measurement of basic characteristics and gain uniformity of a triple GEM detector},
journal = {Nuclear Instruments and Methods in Physics Research Section A: Accelerators, Spectrometers, Detectors and Associated Equipment},
volume = {862},
pages = {25-30},
year = {2017},
issn = {0168-9002},
doi = {https://doi.org/10.1016/j.nima.2017.05.011},
author = {Rajendra Nath Patra and Rama N. Singaraju and Saikat Biswas and Zubayer Ahammed and Tapan K. Nayak and Yogendra P. Viyogi}
}

@article{BACHMANN2002294,
title = {Discharge studies and prevention in the gas electron multiplier (GEM)},
journal = {Nuclear Instruments and Methods in Physics Research Section A: Accelerators, Spectrometers, Detectors and Associated Equipment},
volume = {479},
number = {2},
pages = {294-308},
year = {2002},
issn = {0168-9002},
doi = {https://doi.org/10.1016/S0168-9002(01)00931-7},
author = {S. Bachmann and A. Bressan and M. Capeáns and M. Deutel and S. Kappler and B. Ketzer and A. Polouektov and L. Ropelewski and F. Sauli and E. Schulte and L. Shekhtman and A. Sokolov}
}

@Inbook{Hilke2020,
author="Hilke, H. J.
and Riegler, W.",
editor="Fabjan, Christian W.
and Schopper, Herwig",
title="Gaseous Detectors",
bookTitle="Particle Physics Reference Library: Volume 2: Detectors for Particles and Radiation",
year="2020",
publisher="Springer International Publishing",
address="Cham",
pages="91--136",
isbn="978-3-030-35318-6",
doi="10.1007/978-3-030-35318-6_4",
}

@article{Keller_2020,
doi = {10.1088/1748-0221/15/06/C06004},
year = {2020},
month = {jun},
publisher = {},
volume = {15},
number = {06},
pages = {C06004},
author = {Keller, H. and Hebbeker, T. and Hoepfner, K. and Mocellin, G. and Zaleski, S.},
title = {Influence of hole geometry on gas gain in GEM detectors},
journal = {Journal of Instrumentation}
}

@article{micropattern,
author = {TITOV, MAXIM and ROPELEWSKI, LESZEK},
title = {MICRO-PATTERN GASEOUS DETECTOR TECHNOLOGIES AND RD51 COLLABORATION},
journal = {Modern Physics Letters A},
volume = {28},
number = {13},
pages = {1340022},
year = {2013},
doi = {10.1142/S0217732313400221},
}

@article{BHATTACHARYA201764,
title = {3D simulation of electron and ion transmission of GEM-based detectors},
journal = {Nuclear Instruments and Methods in Physics Research Section A: Accelerators, Spectrometers, Detectors and Associated Equipment},
volume = {870},
pages = {64-72},
year = {2017},
issn = {0168-9002},
doi = {https://doi.org/10.1016/j.nima.2017.06.054},
author = {Purba Bhattacharya and Bedangadas Mohanty and Supratik Mukhopadhyay and Nayana Majumdar and Hugo Natal {da Luz}}
}

@article{Bonivento2002,
  author    = {W. Bonivento and A. Cardini and G. Bencivenni and F. Murtas and D. Pinci},
  title     = {A Complete Simulation of a Triple-GEM Detector},
  journal   = {IEEE Transactions on Nuclear Science},
  volume    = {49},
  number    = {4},
  pages     = {1638--1643},
  year      = {2002},
  doi       = {10.1109/TNS.2002.803785}
}

@article{Kudryavtsev2017,
  author    = {V. N. Kudryavtsev and T. V. Maltsev and L. I. Shekhtman},
  title     = {Study of Spatial Resolution of Coordinate Detectors Based on Gas Electron Multipliers},
  journal   = {Nuclear Instruments and Methods in Physics Research Section A: Accelerators, Spectrometers, Detectors and Associated Equipment},
  volume    = {845},
  pages     = {289--292},
  year      = {2017},
  doi       = {10.1016/j.nima.2016.06.066}
}

@article{Martyanov2014,
  author    = {A. S. Martyanov and N. I. Neustroyev},
  title     = {ANSYS Maxwell Software for Electromagnetic Field Calculations},
  journal   = {Eastern European Scientific Journal},
  number    = {5},
  pages     = {206--210},
  year      = {2014},
  doi       = {10.12851/EESJ201410C05ART03}
}

@article{Jagielski2023,
  author  = {Micha{\l} Jagielski and Karol Malinowski and Maryna Chernyshova},
  title   = {Hybrid Garfield++ simulations of GEM detectors for tokamak plasma radiation monitoring},
  journal = {Fusion Engineering and Design},
  volume  = {195},
  pages   = {113970},
  year    = {2023},
  doi     = {10.1016/j.fusengdes.2023.113970}
}

@incollection{Malhotra2018,
  author    = {Shivali Malhotra and Md. Naimuddin and Ashok Kumar and Mohit Gola and Anshika Bansal and Aashaq Shah},
  title     = {Various Studies with Gas Electron Multiplier (GEM) Detectors},
  booktitle = {Proceedings of the XXII DAE High Energy Physics Symposium},
  series    = {Springer Proceedings in Physics},
  volume    = {203},
  pages     = {227--230},
  publisher = {Springer},
  year      = {2018},
  doi       = {10.1007/978-3-319-73171-1\_22}
}

@article{Tarafdar2022,
  author  = {Sourav Tarafdar and Senta V. Greene and Julia Velkovska and Brandon Blankenship and Michael Z. Reynolds},
  title   = {Reduction of ion backflow using a quadruple GEM detector with various gas mixtures},
  journal = {Nuclear Instruments and Methods in Physics Research Section A},
  volume  = {1046},
  pages   = {167460},
  year    = {2022},
  doi     = {10.1016/j.nima.2022.167460}
}

@article{Sauli2006,
  author  = {Fabio Sauli and Leszek Ropelewski and P. Everaerts},
  title   = {Ion feedback suppression in Time Projection Chambers},
  journal = {Nuclear Instruments and Methods in Physics Research Section A},
  volume  = {560},
  pages   = {269--277},
  year    = {2006},
  doi     = {10.1016/j.nima.2005.12.239}
}

@article{AbiAkl2016,
  author  = {M. Abi-Akl and O. Bouhali and A. Castaneda and Y. Ghorbi and T. Mohammed},
  title   = {Uniformity studies in large area triple-GEM detectors},
  journal = {Nuclear Instruments and Methods in Physics Research Section A},
  volume  = {832},
  pages   = {1--9},
  year    = {2016},
  doi     = {10.1016/j.nima.2016.07.022}
}

@article{Boehmer2013,
  author  = {F. V. B{\"o}hmer and M. Ball and S. D{\o}rheim and C. H{\"o}ppner and G. Lutter and M. Richter and O. Schmidt and M. W{\l}odarczyk},
  title   = {Simulation of space-charge effects in an ungated GEM-based TPC},
  journal = {Nuclear Instruments and Methods in Physics Research Section A},
  volume  = {719},
  pages   = {101--104},
  year    = {2013},
  doi     = {10.1016/j.nima.2013.02.002}
}

@article{Chatterjee2026,
  author  = {Sayak Chatterjee and Supriya Das and Saikat Biswas},
  title   = {The Charging-Up Phenomenon in Gas Electron Multiplier Detectors},
  journal = {Particles},
  volume  = {9},
  number  = {2},
  pages   = {65},
  year    = {2026},
  doi     = {10.3390/particles9020065}
}

@article{Amoroso2024,
  author  = {A. Amoroso and R. Baldini Ferroli and I. Balossino and M. Bertani and D. Bettoni and F. Bianchi and A. Bortone and A. Calcaterra and S. Cerioni and W. Cheng and G. Cibinetto and A. Cotta Ramusino and G. Cotto and F. Cossio and M. Da Rocha Rolo and F. De Mori and M. Destefanis and J. Dong and F. Evangelisti and R. Farinelli and ... and S. Spataro},
  title   = {PARSIFAL: A toolkit for triple-GEM parametrized simulation},
  journal = {Computer Physics Communications},
  volume  = {295},
  pages   = {109000},
  year    = {2024},
  doi     = {10.1016/j.cpc.2023.109000}
}

@article{Bouhali2018,
  author  = {O. Bouhali and A. Sheharyar and T. Mohamed},
  title   = {Accelerating avalanche simulation in gas based charged particle detectors},
  journal = {Nuclear Instruments and Methods in Physics Research Section A},
  volume  = {901},
  pages   = {92--103},
  year    = {2018},
  doi     = {10.1016/j.nima.2018.06.051}
}

@article{Bouhali2022,
  author  = {O. Bouhali and P. Tarjan and H. Feindt and J. Park},
  title   = {Impact of the hole orientation of asymmetric GEM foils on the performance of single and triple GEM detectors},
  journal = {Nuclear Instruments and Methods in Physics Research Section A},
  volume  = {1070},
  pages   = {169257},
  year    = {2022},
  doi     = {10.1016/j.nima.2022.169257}
}

@article{Veenhof1998,
  author  = {R. Veenhof},
  title   = {GARFIELD, recent developments},
  journal = {Nuclear Instruments and Methods in Physics Research Section A},
  volume  = {419},
  pages   = {726--730},
  year    = {1998},
  doi     = {10.1016/S0168-9002(98)00467-9}
}

@article{AlAtoum2020,
  author  = {B. Al Atoum and S. F. Biagi and D. Gonz{\'a}lez-D{\'i}az and L. Lavezzi and M. Caccia and S. Maiolino and M. G. D. Navarra and E. Fragiacomo and L. D. Serrano and E. Nappi},
  title   = {Electron transport in gaseous detectors with a Python-based Monte Carlo simulation code},
  journal = {Computer Physics Communications},
  volume  = {254},
  pages   = {107357},
  year    = {2020},
  doi     = {10.1016/j.cpc.2020.107357}
}

@article{Buzulutskov2007,
  author  = {A. Buzulutskov},
  title   = {Radiation detectors based on gas electron multipliers},
  journal = {Instruments and Experimental Techniques},
  volume  = {50},
  number  = {3},
  pages   = {287--310},
  year    = {2007},
  doi     = {10.1134/S0020441207030080}
}

@article{Bachmann1999,
  author  = {S. Bachmann and A. Bressan and L. Ropelewski and F. Sauli and A. Sharma and H. Eberhardt},
  title   = {Charge amplification and transfer processes in the Gas Electron Multiplier},
  journal = {Nuclear Instruments and Methods in Physics Research Section A},
  volume  = {438},
  pages   = {376--408},
  year    = {1999},
  doi     = {10.1016/S0168-9002(98)01487-1}
}

@article{Bouianov2001,
  author  = {O. Bouianov and P. Bulianov and T. Hemmick and I. Vassiliev},
  title   = {Foil geometry effects on GEM characteristics},
  journal = {Nuclear Instruments and Methods in Physics Research Section A},
  volume  = {458},
  number  = {3},
  pages   = {698--703},
  year    = {2001},
  doi     = {10.1016/S0168-9002(00)00897-4}
}

@article{Bhattacharya2025,
  author  = {Purba Bhattacharya and Promita Roy and Tanay Dey and Jaydeep Datta and Prasant K. Rout and Nayana Majumdar and Supratik Mukhopadhyay},
  title   = {Numerical simulation of charging up, accumulation of space charge and formation of discharges},
  journal = {Nuclear Instruments and Methods in Physics Research Section A},
  volume  = {1075},
  pages   = {170336},
  year    = {2025},
  doi     = {10.1016/j.nima.2025.170336}
}

@article{bachmann1999charge,
  title={Charge amplification and transfer processes in the gas electron multiplier},
  author={Bachmann, S and Bressan, Andrea and Ropelewski, Leszek and Sauli, Fabio and Sharma, A and M{\"o}rmann, D},
  journal={Nuclear Instruments and Methods in Physics Research Section A: Accelerators, Spectrometers, Detectors and Associated Equipment},
  volume={438},
  number={2-3},
  pages={376--408},
  year={1999},
  publisher={Elsevier}
}

@misc{rana2026,
      title={Design Optimization of Triple Gas Electron Multiplier for Superior Gain and Reduced Ion Backflow}, 
      author={Sachin Rana and Md. Kaosor Ali Mondal and Poojan Angiras and Amal Sarkar},
      year={2026},
      eprint={2601.12553},
      archivePrefix={arXiv},
      primaryClass={physics.ins-det},
      url={https://arxiv.org/abs/2601.12553}, 
}
\end{document}